\documentclass[reprint,superscriptaddress,amsmath,amssymb,aps,prb,]{revtex4-2}

\usepackage{graphicx}
\usepackage{siunitx}
\usepackage[version=4]{mhchem}
\usepackage{hyperref}

\begin{document}

\preprint{PRB/LEISonCaSiO}

\title{Towards quantitative Low Energy Ion Scattering on CaSiO$_3$ from Comparison to Multiple-Scattering-Resolved Dynamical Binary Collision Approximation Simulations}

\author{Johannes~Br{\"o}tzner}
\affiliation{%
 TU Wien, Institute of Applied Physics, 1040 Vienna, Austria, EU
}%

\author{Matthias~Kogler}%
\affiliation{%
 TU Wien, Institute of Applied Physics, 1040 Vienna, Austria, EU
}%
\affiliation{%
 Center for Electrochemical Surface Technology, 2700 Wiener Neustadt, Austria, EU
}%

\author{Paul S.~Szabo}%
\affiliation{%
 University of California, Space Sciences Laboratory, Berkeley, CA 94720, United States of America}%

\author{Lukas~Kalchgruber}%
\affiliation{%
 TU Wien, Institute of Applied Physics, 1040 Vienna, Austria, EU
}%
\affiliation{%
 Center for Electrochemical Surface Technology, 2700 Wiener Neustadt, Austria, EU
}%

\author{Andreas~Nenning}
\affiliation{%
 TU Wien, Institute of Chemical Technologies and Analytics, 1060 Vienna, Austria, EU
}%

\author{Andreas~Mutzke}%
\affiliation{%
 Max Planck Institute for Plasma Physics, 17491 Greifswald, Germany, EU}%

\author{Hans~Hofs{\"ass}}%
\affiliation{%
 Universit{\"a}t G{\"o}ttingen, II. Physikalisches Institut, 37077 G{\"o}ttingen, Germany, EU
}%

\author{Markus~Valtiner}%
\affiliation{%
 TU Wien, Institute of Applied Physics, 1040 Vienna, Austria, EU
}%
\affiliation{%
 Center for Electrochemical Surface Technology, 2700 Wiener Neustadt, Austria, EU
}%

\author{Richard~A.~Wilhelm}
\affiliation{%
 TU Wien, Institute of Applied Physics, 1040 Vienna, Austria, EU
}%
 \email{wilhelm@iap.tuwien.ac.at}

\date{\today}

\begin{abstract}
We perform Low Energy Ion Scattering with He ions ranging in energy from \qtyrange{1}{3}{\keV} on CaSiO$_3$ using a commercial electrostatic detector system and determine the charge fraction of scattered ions from comparison with Binary Collision Approximation simulations. The simulations take composition changes due to surface cleaning Xe sputtering into account. Scattered He particles are separated in the simulation into single, dual, and multiple scattering trajectories. We find that the charge fraction of single and dual scattered He is about 10 times higher than the one for multiple collisions. Our results show that the charge fraction of scattered He can be reliably extracted for different scattering regimes and that for a given He incidence energy, the entire scattered He spectrum can be modeled with only few constant parameters. 
\end{abstract}

\keywords{LEIS,BCA,Charge Fraction}

\maketitle

\section{\label{sec:intro}Introduction}

Low Energy Ion Scattering (LEIS) is a powerful method with the ultimate surface sensitivity among all ion beam analysis methods~\cite{prusua_highly_2015, brongersma_low-energy_1978,van_welzenis_low_1999}. Used for both structural analysis by channeling and blocking~\cite{primetzhofer_surface_2007,tromp_new_1984,oconnor_surface_1986} as well as elemental composition analysis~\cite{brongersma_surface_2007,vcushman_low_2016,druce_surface_2014} it is, however, a common misconception that LEIS probes purely the very outermost surface layer~\cite{ter_veen_applications_2009}. Incoming ions can undergo single or multiple scattering events at the surface or in deeper layers that contribute to the final energy spectrum. An additional challenge arises from the fact that the scattered ions may have changed charge states. Some LEIS methods rely on the measurement of the time-of-flight of scattered ions and are therefore not sensitive to changing ion charges~\cite{roth_procedure_2013,markin_origin_2009}, but more commonly, electrostatic LEIS (esaLEIS) is used. The latter has the advantage of typically much higher solid angles especially in commercial devices like the ionTOF Qtac system~\cite{cushman_pictorial_2016}, but suffers from the fact that only charged ions can be detected. 
With a high solid angle, spectra can be acquired in a short time even when detecting only charged scattered ions and consequently dynamic changes in a surface due to, for example, chemical reactions or material deposition can be monitored~\cite{dittmar_application_2017,haunold_ultrahigh_2020}. 

In esaLEIS, a quantification of the elemental composition is limited at best because the charge fraction of scattered ions/atoms is typically unknown and cannot be modeled accurately with current theoretical approaches. As a result, mainly qualitative analysis of dynamical surface changes is performed with esaLEIS~\cite{van_den_berg_investigation_1980}. 

Here we go one step further and show that with the correct simulation toolkit one can in fact extract the charge fraction of scattered He ions even from a complex multi-component target, which also dynamically changes during surface preparation by preceding Xe ion sputtering. The determined charge fraction has only a weak dependence on the dynamic surface changes during Xe sputtering and may therefore serve as a reference for similar (oxide) samples. This implies a possible full quantification when the charge fractions are determined from a sample of known composition first~\cite{gainullin_towards_2018}. Furthermore, our fitting routine uses the entire LEIS energy spectrum and not only a small energy window around the single collision peak. 
In the following, we use CaSiO$_3$ as a target material and analyze the backscattering of He in an energy range of \qtyrange{1}{3}{\keV}. We present a method of LEIS spectra evaluation which will also be applicable to the case of other projectiles and target systems.

\section{\label{sec:methods}Methods}

We used both a commercial LEIS setup and two binary collision approximation (BCA) codes to fit experimental spectra and extract the charge fraction in this procedure. Details are described below.

\subsection{Experimental Setup}

Samples of CaSiO$_3$ were produced by a custom-built Pulsed Laser Deposition setup equipped with a KrF excimer laser (Coherent compex Pro) operating at 5\,Hz. 
At the target 105\,mJ per pulse arrived, which yields a fluence of $3-4$\,J/cm$^2$. The deposition was carried out at a substrate temperature of \qty{300}{\degreeCelsius} in 0.04\,mbar O$_2$ and a target-substrate distance of 6\,cm for 15\,min. We used a natural mineral target~\cite{biber_sputtering_2022} and the substrate is a standard Si(001) wafer piece of \qtyproduct[product-units = single]{1x1}{\cm\squared}. 

The CaSiO$_3$ layer on Si(001) substrate is loaded in an ionTOF Qtac LEIS system with a base pressure in the low $10^{-10}$\,mbar range. The sample is sputtered by Xe$^+$ ions at 1\,keV energy and $60^{\circ}$ incidence angle with respect to the surface normal. An area of $500\times 500$\,$\mu$m$^2$ is irradiated by the Xe ions. After each fluence step of $\Phi_\mathrm{Xe} = 1.0\times 10^{15}$\,Xe/cm$^2$, a LEIS spectrum is taken. In total we apply an Xe fluence of $4.7\times 10^{16}$\,Xe/cm$^2$. Therefore, we track dynamic surface changes during the surface preparation and cleaning procedure.

LEIS spectra are obtained with \qtyrange{1}{3}{\keV} He$^+$ ions under normal incidence. An area of $300\times 300$\,\,$\mu$m$^2$ is probed in the middle of the Xe-sputtered area by the He beam. Scattered He ions are detected by the annular Qtac detector at $145^{\circ} \pm 0.6^{\circ}$ scattering angle. Note that the detector is a toroidal electrostatic analyzer that covers a $360^{\circ}$ azimuthal angle and can only measure charged particles. Ions that neutralize in the process of scattering are not detected. During all our experiments, the energy analyzer is operated in constant pass energy mode. The specific pass energy depends on the He incidence energy and was chosen between 1\,keV and 2\,keV. The analyzer's energy resolution is 1.4\% of the pass energy.

We also used a carbon wafer with lithographic patches of sputter-coated Si, Ni, and Au to determine the single collision peak position for these elements in order to calibrate the spectrometer against the single collision peak position from BCA simulations.

\subsection{Dynamical Sputtering BCA}

We use SDTrimSP 6.09 to extract elemental depth profiles after each Xe sputter fluence we applied in the experiment~\cite{mutzke_sdtrimsp_2019}. We use the dynamic mode of this binary collision approximation code~\cite{hofsass_simulation_2014}. We did not adapt the target density dynamically. The standard surface binding model as well as surface binding energies are used. In the simulation a 40\,nm thick layer with 0.5\,nm discretization is used and a layer intermixing is allowed by the \verb|case_layer_thick=2| option in SDTrimSP.

\subsection{Multiple-Scattering-Resolved BCA}

The elemental depth profiles after Xe sputtering are taken as an input for additional BCA simulations using the IMINTDYN code~\cite{hofsass_low_2023}. 
IMINTDYN is a derivative of SDTrimSP, i.e. using the same BCA description, but including some additional features; for example, to divide particle trajectories between single and multiple scattering. 
With this code we calculate the He scattering yield for a range of $\pm 0.6^{\circ}$ around the mean scattering angle of $145^{\circ}$ using standard ZBL potentials and energy loss functions. 
Note that the exact form of the screened interaction potential and the electronic energy loss model (incl. straggling) can potentially influence the LEIS spectra~\cite{primetzhofer_influence_2011,primetzhofer_leis_2009,primetzhofer_strength_2008,lohmann_trajectory_2023}. 
We further divide the yield into three parts: into scattered particles which scatter once with an angle of $>90^{\circ}$ (single scattering), into scattered particles which scatter once with an angle of $>90^{\circ}$ and once with an angle $\Theta_2$ of $5^{\circ} < \Theta_2 < 90^{\circ}$ (dual scattering), and into scattered particles with all other possible trajectories (multiple scattering). 
The scattering angles $<5^{\circ}$ in a single scattering event are associated with very small energy transfers. 
We also point out that the maximum depth from which ions can still be back-scattered with multiple scattering amounts to $\sim 15$\,nm in our simulations.

\begin{figure}
    \centering
    \includegraphics{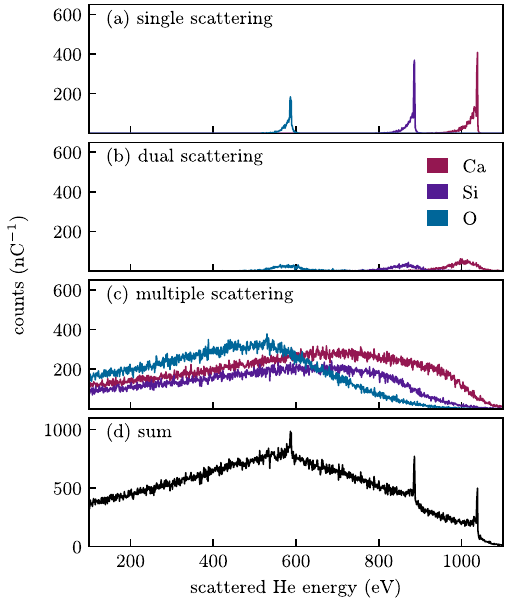}
    \caption{BCA simulation result of 1.5\,keV He scattered for CaSiO$_3$, which was sputtered by $2.5\times 10^{16}$\,Xe/cm$^{2}$ under 60$^{\circ}$ incidence angle in a preceding dynamic BCA simulation. The BCA simulation is decomposed into single scattering (a), dual scattering (b), and multiple scattering (c). The sum of all contributions is shown in (d). The BCA simulation is filtered for the detection angle and range as used in the experiment.}
    \label{fig:bca_example}
\end{figure}

From the IMINTDYN code we additionally resolve the single, dual, and multiple scattering parts of the LEIS spectrum for each target element (see Fig.~\ref{fig:bca_example}). 
For dual and multiple scattering the scattered He is assigned to the element where it first obtains the largest scattering angle.

\section{\label{sec:results}Results}

\begin{figure*}[ht]
    \centering
    \includegraphics{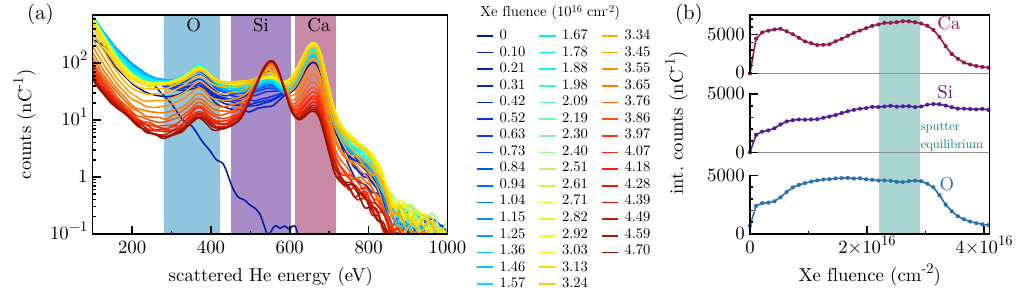}
    \caption{(a) Experimental LEIS spectra using 1\,keV He under normal incidence and 145$^{\circ}$ scattering angle after different sputtering fluences with 1\,keV Xe ions under 60$^{\circ}$ incidence angle. (b) Integrated signal for the colored regions in (a) as function of Xe sputter fluence.}
    \label{fig:experiment}
\end{figure*}

We obtain LEIS spectra after each Xe sputtering step, which can be seen in Fig.~\ref{fig:experiment}(a). 
Initially, the sample was contaminated with light elements (zero Xe fluence, dark blue). After the first Xe sputtering step, the expected multiple scattering background for this sample is recovered in the experiment (blue to cyan, multiple scattering background visible by the signal height in between O and Si). 
At larger fluences (yellow to red), we see a reduced O and Ca signal and an increasing relative contribution of Si. 

In Fig.~\ref{fig:experiment}(b) the integral of the correspondingly colored regions in (a) are shown as a function of the Xe fluence. 
One can see that the signal intensity increases initially due to surface cleaning. 
For $0.5 - 2.2 \times 10^{16}$\,Xe/cm$^2$ a slow variation of the Ca and Si signal can be observed, which we attribute to dynamical surface composition changes due to Xe sputtering. 
For $2.2 - 2.8 \times 10^{16}$\,Xe/cm$^2$ the signal for O, Si, and Ca remain constant, which we assign to the region of sputtering equilibrium, i.e., no more dynamical surface changes occur. 
For even larger fluences the Ca signal as well as the O signal decrease and the Si signal relatively increases, which is understood by sputtering entirely through the PLD-deposited CaSiO$_3$ layer on the Si wafer substrate.

The IMINTDYN simulation provides both element-resolved and multiple-scattering-resolved LEIS spectra. 
One example of a simulated LEIS spectrum after $2.5 \times 10^{16}$\,Xe/cm$^2$ (in the sputter equilibrium) is shown in Fig.~\ref{fig:bca_example} for 1.5\,keV He. 
One can see that single collisions lead to the commonly known sharp peaks in the energy spectrum (a), while dual (b), and multiple (c) scattering lead to increasingly broad spectral components. 
The sum of all components is shown in (d), which would be the LEIS spectrum one would expect if one did not resolve the He trajectories by element or number of scattering events.  

\section{\label{sec:discussion}Discussion}

In order to compare the BCA simulated LEIS spectra with experiment, the experimental LEIS spectra are fit by
\begin{equation}
    f_\mathrm{BCA}^\mathrm{fit}(E) = \sum_{\overset{\substack{X\in\mathrm{ [Ca,\,Si,\,O]}}}{\kappa\in\mathrm{ [s,\,d,\,m]}}}\xi_{\kappa}^X f_{\kappa}^X(E-\delta E^X) , 
    \label{eq:fit}
\end{equation}
through variation of $\xi_\mathrm{s}^{X}$, $\xi_\mathrm{d}^{X}$ and $\xi_\mathrm{m}^{X}$, which are the charge fractions for single, dual, and multiple collisions, respectively, for each element $X$. The $f_\mathrm{s}^{X}(E-\delta E^{X})$, $f_\mathrm{d}^{X}(E-\delta E^{X})$, and $f_\mathrm{m}^{X}(E-\delta E^{X})$ are the partial LEIS spectra from a BCA calculation (cf. Fig.~\ref{fig:bca_example}(a)-(c)), which are additionally convolved with a Gaussian kernel to account for the finite energy resolution of 1.4\% of the pass energy in the experiment. 
The $\delta E^{X}$ are energy shifts to the BCA simulations necessary to obtain a good overall fit (see Appendix~\ref{app:calibration}). 
It is known that the single collision peak is shifted to lower energies in LEIS with respect to the position calculated purely by the scattering kinematics given as a result of the additional electronic energy loss at close collisions~\cite{brongersma_surface_2007}. 
We confirmed that this shift is not a result of the spectrometer energy calibration (see the Appendix~\ref{app:calibration}), but is indeed an element- and He-energy-dependent quantity as expected from the energy loss contributions at close collisions~\cite{prusa_practical_2024}.

\begin{figure}
    \centering
    \includegraphics{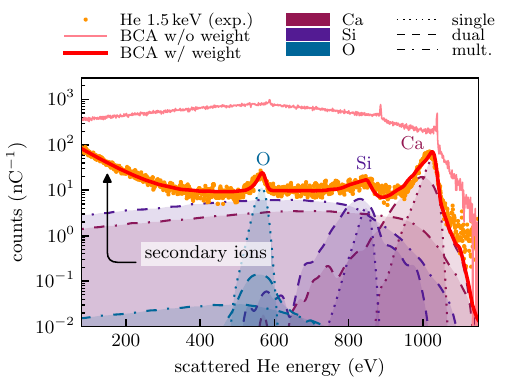}
    \caption{Experimental LEIS spectrum for 1.5\,keV He under normal incidence after $2.5\times 10^{16}$\,Xe/cm$^{2}$, i.e. in the sputter equilibrium regime as well as a BCA simulation for the same conditions. For the BCA simulation a weighted sum of single, dual and multiple scattering for each element was adjusted to fit the experimental spectrum. The fit also contains an exponential background at the low energy part to account for the secondary ions in the spectrum. The weighted BCA spectra (single, dual, and multiple) are convolved with a Gaussian kernel to account for the finite spectrometer resolution in the experiment.}
   \label{fig:fit_example}
\end{figure}

An example of the fit result is shown in Fig.~\ref{fig:fit_example} (red solid line). Also the non-weighted sum (cf. Fig.~\ref{fig:bca_example}(d)) is shown, which is about 2 orders of magnitude larger than the experimental spectrum. 
The average charge fraction for the entire energy range is $\sim 1.9$\%. 
Furthermore, the overall shape is not in agreement with the experiment and only by applying the fit procedure described above can the experimental spectrum be reproduced. 
Note that the experimental spectrum shows an exponential increase to lower energies below 200\,eV, which is commonly attributed to the presence of charged secondary ions in the spectrum and could be filtered from the spectrum using a time-of-flight analysis of the He ions through the electrostatic analyzer~\cite{tellez_new_2014}.
The overall agreement between the BCA fit and the experiment allows the extraction of the charge fractions for the individual scattering events (single, dual, and multiple) for each element separately. 

\begin{figure}
    \centering
    \includegraphics{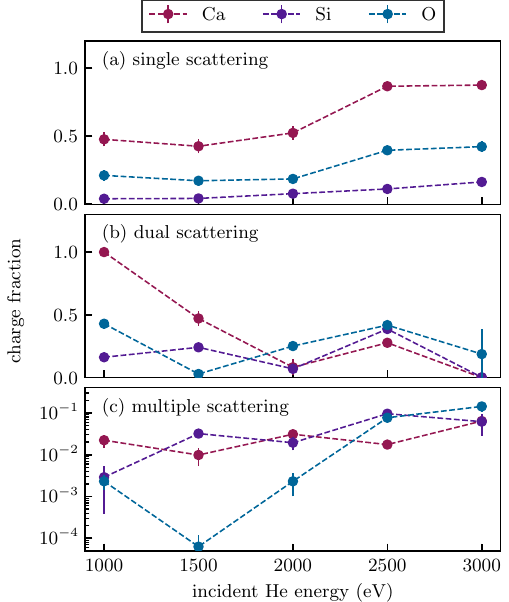}
    \caption{Extracted charge fractions $\xi_{\mathrm{s,d,m}}^{\mathrm{Ca,Si,O}}$ from the fit using Eq.~\ref{eq:fit} for all He energies used in the experiment.}
    \label{fig:charge_fractions}
\end{figure}

Fig.~\ref{fig:charge_fractions} shows the charge fractions $\xi_{\mathrm{s,d,m}}^{\mathrm{Ca,Si,O}}$ extracted from the fit using Eq.~\ref{eq:fit}. 
The charge fraction for Ca is about 50\% for the single collision case below 2.5\,keV He energy and increases to about 80\% for higher He energies. 
The same trend with increasing He energy can be observed for the single collision regime on Si and O while the overall charge fraction is lower than for Ca. 
In particular, the charge fraction for single scattering on Si is below 10\% for all He energies. 
The dual scattering shows at energies below 2\,keV a high charge fraction for Ca between 1 and 0.5. 
For energies of 2\,keV and above all elements show similar charge fractions of about 10\%--40\%. The multiple scattering shows systematically a small charge fraction of below $\sim 10$\% for all energies and elements. 
This can be understood by typically longer trajectories in the solid and more scattering events where the overall likelihood for charge neutralization increases. 
Yet, the multiple scattering contribution extends over the entire energy spectrum (cf. Fig.~\ref{fig:fit_example}) and is not confined to a small energy window as in the case of single and dual scattering. 
Therefore, even a small charge fraction for multiple scattering leads to an overall large spectral weight. 
In particular, the ``background'' of the LEIS spectrum at the single collision peaks is determined mainly by the multiple scattering. 
It is therefore an important finding that the LEIS background can be modeled well by the BCA multiple scattering contribution. 
For this purpose, it is sufficient to consider a charge fraction that, for a given incidence energy and sample species, is constant over a wide range of outgoing projectile energies in the scattered ion spectrum. 
This renders the discussion on LEIS background subtraction potentially obsolete~\cite{avval_calcium_2019,cushman_low_2018, gasteiger_leis_1993}.

The large charge fraction for single and dual scattering in contrast to multiple collisions might result from different active neutralization and re-ionization schemes. 
Single and dual collision trajectories for the lower He ion energies appear close to the surface, i.e., the overall interaction time of the ion with the material is limited at most to a few femtoseconds, so that a substantial fraction of $> 10\%$ of the incoming particles remain in a charged state. 
However, in these two cases relatively strong scattering (i.e., at small interatomic distances) occurs which leads to large deflection angles enabling competing processes such as Auger neutralization and quasi-resonant neutralization~\cite{goebl_low-energy_2015}. 
Multiple scattering, on the other hand, is associated with a comparatively long trajectory inside the material, whereas each scattering event could be considered a soft scattering. 
The interplay and relative importance of impact parameter dependence and interaction time in ion neutralization was recently discussed extensively~\cite{creutzburg_angle-dependent_2021,wilhelm_unraveling_2019} and remains still elusive in heavily asymmetric ion-target combinations like it is the case here. 
At higher He energies the interaction time decreases even further and additional re-ionization channels can open or are enhanced~\cite{goebl_low-energy_2015} which both tend to increase the charge fraction, particularly also for the multiple scattering.

It should be noted that Si, among the three elements probed in our study, is a special case. 
The single scattering charge fraction is much lower than for Ca and O, while the dual and multiple scattering charge fraction is comparable (at least above 1.5\,keV). 
However, the single scattering charge fraction of He at 3\,keV for Si is well in agreement with the literature values of about 14\% determined at 4\,keV~\cite{arezki_angular_1998} and the value of Ca is also well in agreement with previous studies~\cite{sturm_charge_2023}.

The single collision peak of Si shows an energy shift $\delta E^{\mathrm{Si}}\sim 40$\,eV increasing to about 80\,eV at the highest He energy, while these shifts are 10 and 20\,eV for Ca and O, respectively, increasing slowly to 30 and 40\,eV at the highest He energy (see Fig.~\ref{fig:energy_shift} in the Appendix). 
The He-Si scattering shows a larger electronic stopping component in the single collision than the He-Ca and He-O case, which correlates with a larger neutralization for Si than for the other two elements. 
Small impact parameter single scattering for He-Si opens additional charge exchange and correlated energy loss channels~\cite{arnau_molecular-orbital_1995,fano_interpretation_1965}.

\begin{figure}[t!]
    \centering
    \includegraphics{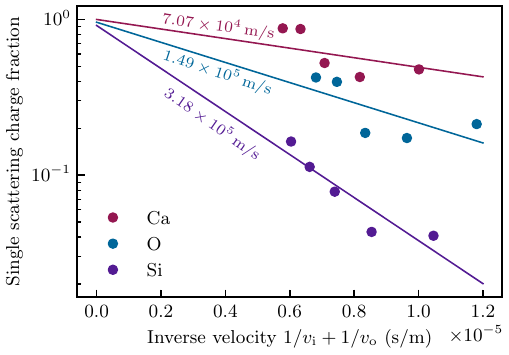}
    \caption{Single scattering charge fractions as a function of inverse ion velocity, considering both ingoing and outgoing He trajectories resolved by sample element. The characteristic velocities, i.e., the slopes in the linear fit of the log of the data, are given in the Figure for each element at the corresponding fit line. }
    \label{fig:chf_inv_vel}
\end{figure}

The herein discussed charged fractions emerge from fitting parameters comparing purely kinematic simulations to experimentally measured ion spectra. 
To investigate whether the numerically obtained weight factors from Eq.~\ref{eq:fit} actually can be physically interpreted as charge fractions, we plotted the single-scattering values over the inverse He ion velocity in Fig.~\ref{fig:chf_inv_vel}, taking into account both the ingoing~($v_\mathrm{in}$) and outgoing trajectories~($v_\mathrm{o}$). 
In the absence of collision-induced neutralization and/or re-ionization, this should yield an exponential behavior~\cite{brongersma_surface_2007, sasaki_neutralization_2002, verbist_calculation_1992}, as is reasonably well fulfilled in our case (note the logarithmic $y$-scale in Fig.~\ref{fig:chf_inv_vel}). 
However, we  point out that the same treatment for the double and multiple scattering contributions is not straightforward, as they extend over a broader energy range of scattered and detected ions, and therefore the velocity for the outgoing trajectory is not well defined anymore. 

Limitations of our method arise when the shapes of the BCA-simulated peaks do not match the experimentally measured ones. 
Since the BCA only accounts for kinematic energy loss (besides a small contribution from an inelastic electronic loss), the physical reasons for such a mismatch could be in energy loss channels not modeled in the BCA. 
One particular example for this is the behavior of the dual scattering charge fraction of 1 for He scattered off Ca at \qty{1}{\keV} incidence energy in Fig.~\ref{fig:charge_fractions}b. 
Here, the residuals between the fit and the experimental spectrum are minimized by retaining the high dual scattering contribution. 
The doubly scattered ions comprise a large fraction of the peak integral, while the singly scattered ones mainly define the high-energy flank of the peak. 
It is inconclusive from our data whether truly the all ions which undergo two scattering events arrive at the detector charged, and especially whether the second collision re-ionizes, or the first medium angle event does not neutralize the projectile. 
For higher incidence energies, the fit quality improves, and the evaluation becomes more robust. 

Similarly, the fit for this case (and all other data points) could potentially be improved if the $\delta E^X$ were allowed to vary with the scattering type for a given element, therefore giving nine rather than three energy shifts. 
However, introducing even more fit parameters would further complicate a meaningful interpretation. 
We thus opt to limit the number of parameters to the current amount, giving a fitting function that is as complex as necessary, yet as simple as possible. 

Once the charge fractions for the given sample are established, they could be used in a quantitative analysis of esaLEIS spectra of a sample with the same constituent properties (i.e., in the same chemical environment), but at different relative proportions. 
The last point is crucial, since the charge fractions depend not only on the scattering partner species alone, but also on the surrounding matrix~\cite{bruckner_neutralization_2020}. 
We also  observed these matrix effects in this study: 
We performed measurements with a pure Si sample for the detector energy calibration as reported in the Appendix~\ref{app:calibration}. 
The single collision charge fractions for Si in this case follow an exponential scaling with inverse velocity as well, but are different from the Si data in Figs.~\ref{fig:charge_fractions} and~\ref{fig:chf_inv_vel}. 
However, we believe that---in an applied context---the data we report here might serve as a reference for the scattering of solar wind He off the alkaline earth silicates like \ce{CaSiO3} or \ce{MgSiO3} on the surfaces of planetary bodies~\cite{bhardwaj_new_2015,lue_artemis_2018,pieters_space_2016,szabo_dynamic_2020}. 

\section{\label{sec:conclusion}Conclusion}

We show that LEIS spectra can be fit by BCA simulations to extract the charge fraction of the scattered He ions using esaLEIS for multi-component oxide target materials. 
The charge fraction depends on the details of the ion trajectory in the surface. 
Dual and multiple scattering can have large contributions to a measured spectrum, indicating that larger depths contribute significantly. 
It is therefore important to properly setup the target surface system in a BCA simulation for LEIS and we do this by dynamically calculating the surface atomic concentration profiles after Xe ion sputtering from separate BCA simulations. 
Some effect of the chemical nature of the scattering partner on the He charge fraction can be inferred from our data as well (Si charge fraction and stopping), while further in-depth analysis on this effect is needed.

Because the determined charge fractions appear to be stable against further Xe sputtering of the surface, we suggest that these charge fractions can be used for similar target systems as a standard. 
It should also be noted that the parameters $\xi^{X}_{\mathrm{m}}$ correlated to the multiple scattering contributions are sufficient to adequately model the experimental spectrum background, despite being constant for a given sample element and He incidence energy. 
This further renders our method practically useful. 

Our approach might help develop esaLEIS into a full quantification method for unknown samples, as long as the charge fraction between single, dual, and multiple scattering remains similar and can be determined for a known standard before. 
The need for a standard sample of known composition due to matrix effects could be potentially eliminated with a sufficiently accurate theory to describe ion charge exchange at surfaces under the relevant conditions.

\begin{acknowledgments}
    This research was funded in whole or in part by the Austrian Science Fund (FWF) [\url{https://doi.org/10.55776/Y1174}, \url{https://doi.org/10.55776/I4914}, \url{https://doi.org/10.55776/P36264}]. 
    For open access purposes, the author has applied a CC BY public copyright license to any author accepted manuscript version arising from this submission. 
    The authors appreciate funding provided EFRE (WST3-F-542638/004-2021) and the State of Lower Austria. 
    J.B. acknowledges financial support by KKKÖ of ÖAW. 
\end{acknowledgments}

\appendix
\section{Fitting procedure}
\label{app:fitting}
In order to fit the BCA spectra to the experimentally obtained ones, we used a non-linear least squares minimization framework as provided by the python LMFIT package~\cite{newville_lmfit_2025}. 
We imposed the following physically motivated constraints on the fitting parameters (i.e., the charge fractions $\xi_{\kappa}^X$ from eq.~\ref{eq:fit}):
\begin{itemize}
    \item Subtraction of an exponential secondary ion background of the form $A\exp(-{E}/{E_\mathrm{decay}})$, where the amplitude $A$ and the characteristic decay energy $E_\mathrm{decay}$ vary with each spectrum and are thus fitting parameters as well.
    \item Single and dual scattering charge fractions are allowed to range from \num{1e-4} to 1, thus fixing a positive minimum contribution and an upper limit corresponding to no charge exchange.
    \item Multiple scattering contributions keep the upper limit of 1. The lower limit is chosen to be zero, ensuring non-negative values, but allowing for smaller contributions than the single and dual scattering.
\end{itemize}

In addition to the constraints described above, a positional shift $\delta E^X$ was introduced for each sample element $X \in $ [Ca, Si, O], effectively shifting the simulated spectral features towards lower energies. 
The necessity to do so was already briefly discussed in Sec.~\ref{sec:discussion}: 
As the kinematic position of the elemental peak at a fixed geometry and scattering angle is idealized and in practice used for elemental identification~\cite{prusa_practical_2024}, in the BCA picture an additional electronic energy loss is accounted for within the target material between collisions. 
The resulting peaks thus appear at somewhat lower energies than calculated purely from the given scattering kinematics. 
The experimental spectra are subject to additional energy losses, e.g. due to charge exchange or re-ionization, not present in purely kinematic or BCA modeling.
Thus, peaks in the experimental spectra are typically observed at energies even lower than in the BCA simulations. 
The $\delta E^X$ in eq.~\ref{eq:fit} are therefore necessary to account for these additional charge-exchange-related energy loss processes and to achieve an optimal fit between experimental and simulated spectra. 

We handled this shift in two ways: 
First, we used a peak finder to algorithmically determine the shift between the experimental single-collision peak and the BCA-simulated single-scattering peak for every target element. 
Second, we allowed this shift to vary within $\pm 5\%$ of this detected value. 
The approach of matching spectral features (peaks) relies heavily on a rather arbitrary definition of peak positions (actual position vs. 60\%/40\% partition vs. any similar partition), we allowed the determined shift to be varied by $\pm 5\%$ in the second step. 
Within this interval, it is assured that, on the one hand, the fitting algorithm can find a better match between the spectra and, on the other hand, that the spectra are not shifted too much. 
This means that most of the physical information is still carried by the single-collision peak. 

Moreover, we observed that with pass energies higher than \qty{1.5}{\keV}, the experimental peaks appeared to be sharper than the smoothened BCA spectra. 
This could potentially mean that, at these energies, the resolution of the Qtac detector might actually be slightly better than $1.4\%$ of the pass energy as specified by the manufacturer. 
To gauge the influence of this effect on the extracted charge fractions, we also performed the fitting routine with Gaussian filters where the FWHM was fixed to $1.4\%$ of $E_\mathrm{pass}=1.5$\,keV where pass energies exceeded this value. 

\begin{figure*}
    \centering
    \includegraphics{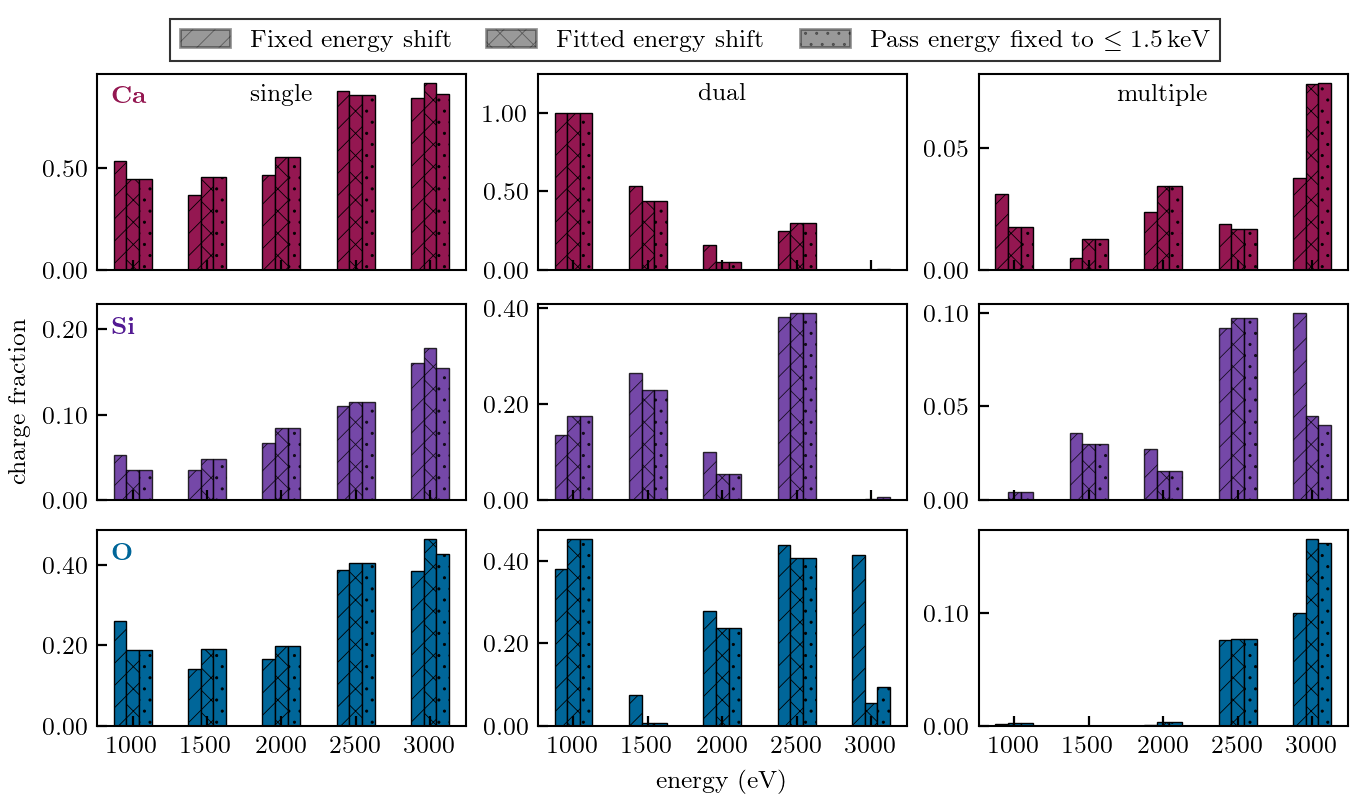}
    \caption{Extracted charge fractions for single, dual, and multiple scattering on Ca, Si, and O for all He energies using three varying fitting routines. The energy shifts $\delta E^X$ were kept constant (bars with diagonal hatches), allowed to vary within $\pm 5\%$ of the constant value (crossed hatches), as well as results for an approach where simulated detector resolution was kept high for high pass energy cases (dotted bars).}
    \label{fig:all_charge_fractions}
\end{figure*}

The charge fractions resulting from these three fitting procedures: (1) a fixed energy shift determined from peak positions, (2) a fitting of this energy shift, as well as (3) the artificially fixed detector resolution are given in Fig.~\ref{fig:all_charge_fractions}. 
In this figure, the charge fractions are resolved by target element (rows) as well as scattering event type (columns). 
For a given species and scattering type, the standard deviation between the fitting routines defines the error bars in Fig.~\ref{fig:charge_fractions}.

It should be noted from Fig.~\ref{fig:all_charge_fractions} that the charge fractions are consistent between all three possible fitting procedures/constraints with some ambiguity in the multiple scattering regime at 3\,keV He energy.

\section{Energy calibration}
\label{app:calibration}

As the energy shift described above is necessary to achieve an optimal fit between experiment and simulation, we carried out LEIS experiments on a calibration sample of lithographically deposited patches of Si, Ni and Au to ensure that the observed shifts are indeed physical rather than artifacts of an ill-calibrated detector. 
We performed experiments using He of energies ranging from \qtyrange[]{1}{5}{\keV} in increments of \qty{500}{\eV} as well as the corresponding IMINTDYN simulations. 
Just as above, we extracted the peak positions from both spectra types. 
While no direct perfect match between the measured and simulated peak positions is expected due to energy loss contributions that are not modeled in BCA, a well-calibrated detector should exhibit a linear correlation between the two. 

\begin{figure}[th]
    \centering
    \includegraphics{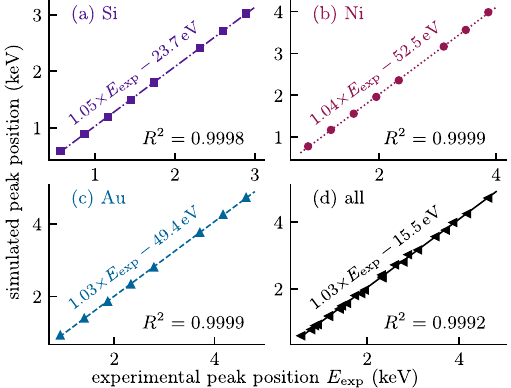}
    \caption{Energy calibration between BCA-simulated single collision peak positions and experimentally-determined single collision peak positions for (a) Si, (b) Ni, (c) Au, and (d) taking all peaks into account.}
    \label{fig:energy_calibration}
\end{figure}

Figs.~\ref{fig:energy_calibration}(a) to \ref{fig:energy_calibration}(c) show the relations of peak positions for Si, Ni and Au, respectively, given by a linear fit. 
It is apparent that in all cases, the relation can be confidently assumed to be linear as indicated by the $R^2$ coefficients of determination very close to 1. 
Furthermore, panel \ref{fig:energy_calibration}(d) shows the calibration if all data points are taken into account, regardless of the sample species. 

A comparison between the elementally resolved and the overall linear fits yields two main points. 
First, the overall relation over all sample species has the slope and the intercept closest to unity and zero, respectively. 
This indicates that in the global picture, the detector can be viewed as well calibrated and that the observed energy shifts are not instrumentation artifacts. 
Second, when considered individually, the fitting parameters vary between the sample species. 
The latter implies that the energy shifts are species-dependent, thus corroborating the approach to determine an elementally resolved energy shift $\delta E^X$ in eq.~\ref{eq:fit} rather than trying to find one global energy calibration.

\begin{figure}[htbp]
    \centering
    \includegraphics{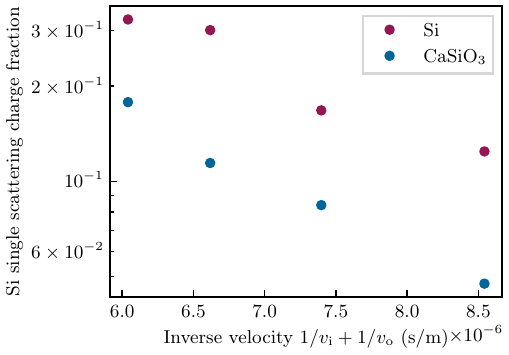}
    \caption{Si single scattering charge compared across two sample matrices. Matrix effects are clearly present between charge fractions extracted from a pure sample (red circles) compared the ones obtained for the Ca silicate (blue circles). }
    \label{fig:chf_comparison_Si}
\end{figure}

On a side note, the calibration measurements with the Si sample allow for a comparison of the charge fractions of the same element, but embedded into different sample matrices. 
We therefore give the extracted single scattering charge fractions for the pure Si case in Fig.~\ref{fig:chf_comparison_Si}. 
Both data sets follow the exponential scaling similarly well, but are significantly distinct. 
As described in section~\ref{sec:discussion}, these matrix effects limit the application reference charge fractions for sample with different chemical environments. 

\section{Energy shift and electronic stopping}

\begin{figure}[htbp]
    \centering
    \includegraphics{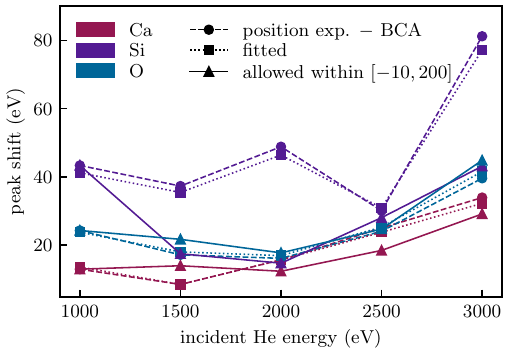}
    \caption{Determined single collision peak energy shifts for different incident He energies on the different elements Ca, Si, and O. We used three different approaches to determine this shift and find consistent values}
    \label{fig:energy_shift}
\end{figure}

The charge-exchange-related energy shifts are determined by the difference between experimental peak positions and BCA-simulated single-collision peak positions, see Fig.~\ref{fig:energy_shift}. We also determined this peak shift independently by allowing the fitting algorithm to freely fit the $\delta E^X$ shifts. We did the same, but constrained the possible peak shifts between [-10,200]\,eV. In all cases we find an increase of this charge-exchange-related energy loss with increasing He energy. Furthermore, the values are between 5 and 25\,eV for Ca, 18 and 35\,eV for O, and 30 to 80\,eV for Si. Si shows consistently the highest peak shifts (and lowest single collision charge fraction).


\begin{thebibliography}{45}%
\makeatletter
\providecommand \@ifxundefined [1]{%
 \@ifx{#1\undefined}
}%
\providecommand \@ifnum [1]{%
 \ifnum #1\expandafter \@firstoftwo
 \else \expandafter \@secondoftwo
 \fi
}%
\providecommand \@ifx [1]{%
 \ifx #1\expandafter \@firstoftwo
 \else \expandafter \@secondoftwo
 \fi
}%
\providecommand \natexlab [1]{#1}%
\providecommand \enquote  [1]{``#1''}%
\providecommand \bibnamefont  [1]{#1}%
\providecommand \bibfnamefont [1]{#1}%
\providecommand \citenamefont [1]{#1}%
\providecommand \href@noop [0]{\@secondoftwo}%
\providecommand \href [0]{\begingroup \@sanitize@url \@href}%
\providecommand \@href[1]{\@@startlink{#1}\@@href}%
\providecommand \@@href[1]{\endgroup#1\@@endlink}%
\providecommand \@sanitize@url [0]{\catcode `\\12\catcode `\$12\catcode
  `\&12\catcode `\#12\catcode `\^12\catcode `\_12\catcode `\%12\relax}%
\providecommand \@@startlink[1]{}%
\providecommand \@@endlink[0]{}%
\providecommand \url  [0]{\begingroup\@sanitize@url \@url }%
\providecommand \@url [1]{\endgroup\@href {#1}{\urlprefix }}%
\providecommand \urlprefix  [0]{URL }%
\providecommand \Eprint [0]{\href }%
\providecommand \doibase [0]{https://doi.org/}%
\providecommand \selectlanguage [0]{\@gobble}%
\providecommand \bibinfo  [0]{\@secondoftwo}%
\providecommand \bibfield  [0]{\@secondoftwo}%
\providecommand \translation [1]{[#1]}%
\providecommand \BibitemOpen [0]{}%
\providecommand \bibitemStop [0]{}%
\providecommand \bibitemNoStop [0]{.\EOS\space}%
\providecommand \EOS [0]{\spacefactor3000\relax}%
\providecommand \BibitemShut  [1]{\csname bibitem#1\endcsname}%
\let\auto@bib@innerbib\@empty
%</preamble>
\bibitem [{\citenamefont {Průša}\ \emph {et~al.}(2015)\citenamefont {Průša},
  \citenamefont {Prochazka}, \citenamefont {Bábor}, \citenamefont {Sikola},
  \citenamefont {Ter~Veen}, \citenamefont {Fartmann}, \citenamefont {Grehl},
  \citenamefont {Brüner}, \citenamefont {Roth}, \citenamefont {Bauer},\ and\
  \citenamefont {Brongersma}}]{prusua_highly_2015}%
  \BibitemOpen
  \bibfield  {author} {\bibinfo {author} {\bibfnamefont {S.}~\bibnamefont
  {Průša}}, \bibinfo {author} {\bibfnamefont {P.}~\bibnamefont {Prochazka}},
  \bibinfo {author} {\bibfnamefont {P.}~\bibnamefont {Bábor}}, \bibinfo
  {author} {\bibfnamefont {T.}~\bibnamefont {Sikola}}, \bibinfo {author}
  {\bibfnamefont {R.}~\bibnamefont {Ter~Veen}}, \bibinfo {author}
  {\bibfnamefont {M.}~\bibnamefont {Fartmann}}, \bibinfo {author}
  {\bibfnamefont {T.}~\bibnamefont {Grehl}}, \bibinfo {author} {\bibfnamefont
  {P.}~\bibnamefont {Brüner}}, \bibinfo {author} {\bibfnamefont
  {D.}~\bibnamefont {Roth}}, \bibinfo {author} {\bibfnamefont {P.}~\bibnamefont
  {Bauer}},\ and\ \bibinfo {author} {\bibfnamefont {H.~H.}\ \bibnamefont
  {Brongersma}},\ }\bibfield  {title} {\bibinfo {title} {Highly {Sensitive}
  {Detection} of {Surface} and {Intercalated} {Impurities} in {Graphene} by
  {LEIS}},\ }\href {https://doi.org/10.1021/acs.langmuir.5b01935} {\bibfield
  {journal} {\bibinfo  {journal} {Langmuir}\ }\textbf {\bibinfo {volume}
  {31}},\ \bibinfo {pages} {9628} (\bibinfo {year} {2015})}\BibitemShut
  {NoStop}%
\bibitem [{\citenamefont {Brongersma}\ and\ \citenamefont
  {Buck}(1978)}]{brongersma_low-energy_1978}%
  \BibitemOpen
  \bibfield  {author} {\bibinfo {author} {\bibfnamefont {H.~H.}\ \bibnamefont
  {Brongersma}}\ and\ \bibinfo {author} {\bibfnamefont {T.~M.}\ \bibnamefont
  {Buck}},\ }\bibfield  {title} {\bibinfo {title} {Low-energy ion scattering
  ({LEIS}) for composition and structure analysis of the outer surface},\
  }\href {https://doi.org/10.1016/0029-554X(78)90929-1} {\bibfield  {journal}
  {\bibinfo  {journal} {Nuclear Instruments and Methods}\ }\textbf {\bibinfo
  {volume} {149}},\ \bibinfo {pages} {569} (\bibinfo {year}
  {1978})}\BibitemShut {NoStop}%
\bibitem [{\citenamefont {van Welzenis}(1999)}]{van_welzenis_low_1999}%
  \BibitemOpen
  \bibfield  {author} {\bibinfo {author} {\bibfnamefont {R.~G.}\ \bibnamefont
  {van Welzenis}},\ }\bibfield  {title} {\bibinfo {title} {Low energy ion
  scattering ({LEIS}) applied to ionic materials},\ }\href
  {https://doi.org/10.1007/BF02375897} {\bibfield  {journal} {\bibinfo
  {journal} {Ionics}\ }\textbf {\bibinfo {volume} {5}},\ \bibinfo {pages} {13}
  (\bibinfo {year} {1999})}\BibitemShut {NoStop}%
\bibitem [{\citenamefont {Primetzhofer}\ \emph {et~al.}(2007)\citenamefont
  {Primetzhofer}, \citenamefont {Markin}, \citenamefont {Kolarova},
  \citenamefont {Draxler}, \citenamefont {Beikler}, \citenamefont {Taglauer},\
  and\ \citenamefont {Bauer}}]{primetzhofer_surface_2007}%
  \BibitemOpen
  \bibfield  {author} {\bibinfo {author} {\bibfnamefont {D.}~\bibnamefont
  {Primetzhofer}}, \bibinfo {author} {\bibfnamefont {S.}~\bibnamefont
  {Markin}}, \bibinfo {author} {\bibfnamefont {R.}~\bibnamefont {Kolarova}},
  \bibinfo {author} {\bibfnamefont {M.}~\bibnamefont {Draxler}}, \bibinfo
  {author} {\bibfnamefont {R.}~\bibnamefont {Beikler}}, \bibinfo {author}
  {\bibfnamefont {E.}~\bibnamefont {Taglauer}},\ and\ \bibinfo {author}
  {\bibfnamefont {P.}~\bibnamefont {Bauer}},\ }\bibfield  {title} {\bibinfo
  {title} {On the surface sensitivity of angular scans in {LEIS}},\ }\href
  {https://doi.org/10.1016/j.nimb.2006.12.176} {\bibfield  {journal} {\bibinfo
  {journal} {Nuclear Instruments and Methods in Physics Research Section B:
  Beam Interactions with Materials and Atoms}\ }\textbf {\bibinfo {volume}
  {258}},\ \bibinfo {pages} {36} (\bibinfo {year} {2007})}\BibitemShut
  {NoStop}%
\bibitem [{\citenamefont {Tromp}\ \emph {et~al.}(1984)\citenamefont {Tromp},
  \citenamefont {Kersten}, \citenamefont {Granneman}, \citenamefont {Saris},
  \citenamefont {Koudijs},\ and\ \citenamefont {Kilsdonk}}]{tromp_new_1984}%
  \BibitemOpen
  \bibfield  {author} {\bibinfo {author} {\bibfnamefont {R.}~\bibnamefont
  {Tromp}}, \bibinfo {author} {\bibfnamefont {H.}~\bibnamefont {Kersten}},
  \bibinfo {author} {\bibfnamefont {E.}~\bibnamefont {Granneman}}, \bibinfo
  {author} {\bibfnamefont {F.}~\bibnamefont {Saris}}, \bibinfo {author}
  {\bibfnamefont {R.}~\bibnamefont {Koudijs}},\ and\ \bibinfo {author}
  {\bibfnamefont {W.}~\bibnamefont {Kilsdonk}},\ }\bibfield  {title} {\bibinfo
  {title} {A new {UHV} system for channeling blocking analysis of solid
  surfaces and interfaces},\ }\href
  {https://doi.org/10.1016/0168-583X(84)90055-7} {\bibfield  {journal}
  {\bibinfo  {journal} {Nuclear Instruments and Methods in Physics Research
  Section B: Beam Interactions with Materials and Atoms}\ }\textbf {\bibinfo
  {volume} {4}},\ \bibinfo {pages} {155} (\bibinfo {year} {1984})}\BibitemShut
  {NoStop}%
\bibitem [{\citenamefont {O'connor}\ \emph {et~al.}(1986)\citenamefont
  {O'connor}, \citenamefont {Macdonald}, \citenamefont {Eckstein},\ and\
  \citenamefont {Higginbottom}}]{oconnor_surface_1986}%
  \BibitemOpen
  \bibfield  {author} {\bibinfo {author} {\bibfnamefont {D.}~\bibnamefont
  {O'connor}}, \bibinfo {author} {\bibfnamefont {R.}~\bibnamefont {Macdonald}},
  \bibinfo {author} {\bibfnamefont {W.}~\bibnamefont {Eckstein}},\ and\
  \bibinfo {author} {\bibfnamefont {P.}~\bibnamefont {Higginbottom}},\
  }\bibfield  {title} {\bibinfo {title} {Surface structure analysis using low
  energy scattered and recoiling ions},\ }\href
  {https://doi.org/10.1016/0168-583X(86)90509-4} {\bibfield  {journal}
  {\bibinfo  {journal} {Nuclear Instruments and Methods in Physics Research
  Section B: Beam Interactions with Materials and Atoms}\ }\textbf {\bibinfo
  {volume} {13}},\ \bibinfo {pages} {235} (\bibinfo {year} {1986})}\BibitemShut
  {NoStop}%
\bibitem [{\citenamefont {Brongersma}\ \emph {et~al.}(2007)\citenamefont
  {Brongersma}, \citenamefont {Draxler}, \citenamefont {Deridder},\ and\
  \citenamefont {Bauer}}]{brongersma_surface_2007}%
  \BibitemOpen
  \bibfield  {author} {\bibinfo {author} {\bibfnamefont {H.}~\bibnamefont
  {Brongersma}}, \bibinfo {author} {\bibfnamefont {M.}~\bibnamefont {Draxler}},
  \bibinfo {author} {\bibfnamefont {M.}~\bibnamefont {Deridder}},\ and\
  \bibinfo {author} {\bibfnamefont {P.}~\bibnamefont {Bauer}},\ }\bibfield
  {title} {\bibinfo {title} {Surface composition analysis by low-energy ion
  scattering},\ }\href {https://doi.org/10.1016/j.surfrep.2006.12.002}
  {\bibfield  {journal} {\bibinfo  {journal} {Surface Science Reports}\
  }\textbf {\bibinfo {volume} {62}},\ \bibinfo {pages} {63} (\bibinfo {year}
  {2007})}\BibitemShut {NoStop}%
\bibitem [{\citenamefont {V. Cushman}\ \emph {et~al.}(2016)\citenamefont
  {V. Cushman}, \citenamefont {Brüner}, \citenamefont {Zakel}, \citenamefont
  {H. Major}, \citenamefont {M. Lunt}, \citenamefont {J. Smith},
  \citenamefont {Grehl},\ and\ \citenamefont
  {R. Linford}}]{vcushman_low_2016}%
  \BibitemOpen
  \bibfield  {author} {\bibinfo {author} {\bibfnamefont {C.}~\bibnamefont
  {V. Cushman}}, \bibinfo {author} {\bibfnamefont {P.}~\bibnamefont
  {Brüner}}, \bibinfo {author} {\bibfnamefont {J.}~\bibnamefont {Zakel}},
  \bibinfo {author} {\bibfnamefont {G.}~\bibnamefont {H. Major}}, \bibinfo
  {author} {\bibfnamefont {B.}~\bibnamefont {M. Lunt}}, \bibinfo {author}
  {\bibfnamefont {N.}~\bibnamefont {J. Smith}}, \bibinfo {author}
  {\bibfnamefont {T.}~\bibnamefont {Grehl}},\ and\ \bibinfo {author}
  {\bibfnamefont {M.}~\bibnamefont {R. Linford}},\ }\bibfield  {title}
  {\bibinfo {title} {Low energy ion scattering ({LEIS}). {A} practical
  introduction to its theory, instrumentation, and applications},\ }\href
  {https://doi.org/10.1039/C6AY00765A} {\bibfield  {journal} {\bibinfo
  {journal} {Analytical Methods}\ }\textbf {\bibinfo {volume} {8}},\ \bibinfo
  {pages} {3419} (\bibinfo {year} {2016})},\ \bibinfo {note} {publisher: Royal
  Society of Chemistry}\BibitemShut {NoStop}%
\bibitem [{\citenamefont {Druce}\ \emph {et~al.}(2014)\citenamefont {Druce},
  \citenamefont {Ishihara},\ and\ \citenamefont {Kilner}}]{druce_surface_2014}%
  \BibitemOpen
  \bibfield  {author} {\bibinfo {author} {\bibfnamefont {J.}~\bibnamefont
  {Druce}}, \bibinfo {author} {\bibfnamefont {T.}~\bibnamefont {Ishihara}},\
  and\ \bibinfo {author} {\bibfnamefont {J.}~\bibnamefont {Kilner}},\
  }\bibfield  {title} {\bibinfo {title} {Surface composition of perovskite-type
  materials studied by {Low} {Energy} {Ion} {Scattering} ({LEIS})},\ }\href
  {https://doi.org/10.1016/j.ssi.2013.09.010} {\bibfield  {journal} {\bibinfo
  {journal} {Solid State Ionics}\ }\bibinfo {series} {Solid {State} {Ionics} 19
  {Proceedings} of the 19th {International} {Conference} on {Solid} {State}
  {Ionics}},\ \textbf {\bibinfo {volume} {262}},\ \bibinfo {pages} {893}
  (\bibinfo {year} {2014})}\BibitemShut {NoStop}%
\bibitem [{\citenamefont {Ter~Veen}\ \emph {et~al.}(2009)\citenamefont
  {Ter~Veen}, \citenamefont {Kim}, \citenamefont {Wachs},\ and\ \citenamefont
  {Brongersma}}]{ter_veen_applications_2009}%
  \BibitemOpen
  \bibfield  {author} {\bibinfo {author} {\bibfnamefont {H.}~\bibnamefont
  {Ter~Veen}}, \bibinfo {author} {\bibfnamefont {T.}~\bibnamefont {Kim}},
  \bibinfo {author} {\bibfnamefont {I.}~\bibnamefont {Wachs}},\ and\ \bibinfo
  {author} {\bibfnamefont {H.}~\bibnamefont {Brongersma}},\ }\bibfield  {title}
  {\bibinfo {title} {Applications of {High} {Sensitivity}-{Low} {Energy} {Ion}
  {Scattering} ({HS}-{LEIS}) in heterogeneous catalysis},\ }\href
  {https://doi.org/10.1016/j.cattod.2008.10.012} {\bibfield  {journal}
  {\bibinfo  {journal} {Catalysis Today}\ }\textbf {\bibinfo {volume} {140}},\
  \bibinfo {pages} {197} (\bibinfo {year} {2009})}\BibitemShut {NoStop}%
\bibitem [{\citenamefont {Roth}\ \emph {et~al.}(2013)\citenamefont {Roth},
  \citenamefont {Goebl}, \citenamefont {Primetzhofer},\ and\ \citenamefont
  {Bauer}}]{roth_procedure_2013}%
  \BibitemOpen
  \bibfield  {author} {\bibinfo {author} {\bibfnamefont {D.}~\bibnamefont
  {Roth}}, \bibinfo {author} {\bibfnamefont {D.}~\bibnamefont {Goebl}},
  \bibinfo {author} {\bibfnamefont {D.}~\bibnamefont {Primetzhofer}},\ and\
  \bibinfo {author} {\bibfnamefont {P.}~\bibnamefont {Bauer}},\ }\bibfield
  {title} {\bibinfo {title} {A procedure to determine electronic energy loss
  from relative measurements with {TOF}-{LEIS}},\ }\href
  {https://doi.org/10.1016/j.nimb.2012.12.094} {\bibfield  {journal} {\bibinfo
  {journal} {Nuclear Instruments and Methods in Physics Research Section B:
  Beam Interactions with Materials and Atoms}\ }\textbf {\bibinfo {volume}
  {317}},\ \bibinfo {pages} {61} (\bibinfo {year} {2013})}\BibitemShut
  {NoStop}%
\bibitem [{\citenamefont {Markin}\ \emph {et~al.}(2009)\citenamefont {Markin},
  \citenamefont {Primetzhofer},\ and\ \citenamefont
  {Bauer}}]{markin_origin_2009}%
  \BibitemOpen
  \bibfield  {author} {\bibinfo {author} {\bibfnamefont {S.}~\bibnamefont
  {Markin}}, \bibinfo {author} {\bibfnamefont {D.}~\bibnamefont
  {Primetzhofer}},\ and\ \bibinfo {author} {\bibfnamefont {P.}~\bibnamefont
  {Bauer}},\ }\bibfield  {title} {\bibinfo {title} {On the origin of the {LEIS}
  signal in {TOF}- and in {ESA}-{LEIS}},\ }\href
  {https://doi.org/10.1016/j.nimb.2008.11.022} {\bibfield  {journal} {\bibinfo
  {journal} {Nuclear Instruments and Methods in Physics Research Section B:
  Beam Interactions with Materials and Atoms}\ }\textbf {\bibinfo {volume}
  {267}},\ \bibinfo {pages} {634} (\bibinfo {year} {2009})}\BibitemShut
  {NoStop}%
\bibitem [{\citenamefont {Cushman}\ \emph {et~al.}(2016)\citenamefont
  {Cushman}, \citenamefont {Brüner}, \citenamefont {Zakel}, \citenamefont
  {Major}, \citenamefont {Lunt}, \citenamefont {Grehl}, \citenamefont {Smith},\
  and\ \citenamefont {Linford}}]{cushman_pictorial_2016}%
  \BibitemOpen
  \bibfield  {author} {\bibinfo {author} {\bibfnamefont {C.~V.}\ \bibnamefont
  {Cushman}}, \bibinfo {author} {\bibfnamefont {P.}~\bibnamefont {Brüner}},
  \bibinfo {author} {\bibfnamefont {J.}~\bibnamefont {Zakel}}, \bibinfo
  {author} {\bibfnamefont {G.}~\bibnamefont {Major}}, \bibinfo {author}
  {\bibfnamefont {B.~M.}\ \bibnamefont {Lunt}}, \bibinfo {author}
  {\bibfnamefont {T.}~\bibnamefont {Grehl}}, \bibinfo {author} {\bibfnamefont
  {N.~J.}\ \bibnamefont {Smith}},\ and\ \bibinfo {author} {\bibfnamefont
  {M.~R.}\ \bibnamefont {Linford}},\ }\bibfield  {title} {\bibinfo {title} {A
  pictorial view of {LEIS} and {ToF}-{SIMS} instrumentation.},\ }\href@noop {}
  {\bibfield  {journal} {\bibinfo  {journal} {Vacuum Technology \& Coating}\ ,\
  \bibinfo {pages} {27}} (\bibinfo {year} {2016})}\BibitemShut {NoStop}%
\bibitem [{\citenamefont {Dittmar}\ \emph {et~al.}(2017)\citenamefont
  {Dittmar}, \citenamefont {Triyoso}, \citenamefont {Erben}, \citenamefont
  {Metzger}, \citenamefont {Binder}, \citenamefont {Brongersma}, \citenamefont
  {Weisheit},\ and\ \citenamefont {Engelmann}}]{dittmar_application_2017}%
  \BibitemOpen
  \bibfield  {author} {\bibinfo {author} {\bibfnamefont {K.}~\bibnamefont
  {Dittmar}}, \bibinfo {author} {\bibfnamefont {D.~H.}\ \bibnamefont
  {Triyoso}}, \bibinfo {author} {\bibfnamefont {E.}~\bibnamefont {Erben}},
  \bibinfo {author} {\bibfnamefont {J.}~\bibnamefont {Metzger}}, \bibinfo
  {author} {\bibfnamefont {R.}~\bibnamefont {Binder}}, \bibinfo {author}
  {\bibfnamefont {H.~H.}\ \bibnamefont {Brongersma}}, \bibinfo {author}
  {\bibfnamefont {M.}~\bibnamefont {Weisheit}},\ and\ \bibinfo {author}
  {\bibfnamefont {H.-J.}\ \bibnamefont {Engelmann}},\ }\bibfield  {title}
  {\bibinfo {title} {The application of low energy ion scattering spectroscopy
  ({LEIS}) in sub 28-nm {CMOS} technology},\ }\href
  {https://doi.org/10.1002/sia.6312} {\bibfield  {journal} {\bibinfo  {journal}
  {Surface and Interface Analysis}\ }\textbf {\bibinfo {volume} {49}},\
  \bibinfo {pages} {1175} (\bibinfo {year} {2017})}\BibitemShut {NoStop}%
\bibitem [{\citenamefont {Haunold}\ \emph {et~al.}(2020)\citenamefont
  {Haunold}, \citenamefont {Rameshan}, \citenamefont {Bukhtiyarov},\ and\
  \citenamefont {Rupprechter}}]{haunold_ultrahigh_2020}%
  \BibitemOpen
  \bibfield  {author} {\bibinfo {author} {\bibfnamefont {T.}~\bibnamefont
  {Haunold}}, \bibinfo {author} {\bibfnamefont {C.}~\bibnamefont {Rameshan}},
  \bibinfo {author} {\bibfnamefont {A.~V.}\ \bibnamefont {Bukhtiyarov}},\ and\
  \bibinfo {author} {\bibfnamefont {G.}~\bibnamefont {Rupprechter}},\
  }\bibfield  {title} {\bibinfo {title} {An ultrahigh vacuum-compatible
  reaction cell for model catalysis under atmospheric pressure flow
  conditions},\ }\href {https://doi.org/10.1063/5.0026171} {\bibfield
  {journal} {\bibinfo  {journal} {Review of Scientific Instruments}\ }\textbf
  {\bibinfo {volume} {91}},\ \bibinfo {pages} {125101} (\bibinfo {year}
  {2020})}\BibitemShut {NoStop}%
\bibitem [{\citenamefont {Van Den~Berg}\ \emph {et~al.}(1980)\citenamefont {Van
  Den~Berg}, \citenamefont {Verheij},\ and\ \citenamefont
  {Armour}}]{van_den_berg_investigation_1980}%
  \BibitemOpen
  \bibfield  {author} {\bibinfo {author} {\bibfnamefont {J.}~\bibnamefont {Van
  Den~Berg}}, \bibinfo {author} {\bibfnamefont {L.}~\bibnamefont {Verheij}},\
  and\ \bibinfo {author} {\bibfnamefont {D.}~\bibnamefont {Armour}},\
  }\bibfield  {title} {\bibinfo {title} {An investigation of the kinetics of
  structural changes during the early oxidation stages of a {Ni}(100) surface
  using low energy ion scattering ({LEIS})},\ }\href
  {https://doi.org/10.1016/0039-6028(80)90081-3} {\bibfield  {journal}
  {\bibinfo  {journal} {Surface Science}\ }\textbf {\bibinfo {volume} {91}},\
  \bibinfo {pages} {218} (\bibinfo {year} {1980})}\BibitemShut {NoStop}%
\bibitem [{\citenamefont {Gainullin}(2018)}]{gainullin_towards_2018}%
  \BibitemOpen
  \bibfield  {author} {\bibinfo {author} {\bibfnamefont {I.}~\bibnamefont
  {Gainullin}},\ }\bibfield  {title} {\bibinfo {title} {Towards quantitative
  {LEIS} with alkali metal ions},\ }\href
  {https://doi.org/10.1016/j.susc.2018.08.007} {\bibfield  {journal} {\bibinfo
  {journal} {Surface Science}\ }\textbf {\bibinfo {volume} {677}},\ \bibinfo
  {pages} {324} (\bibinfo {year} {2018})}\BibitemShut {NoStop}%
\bibitem [{\citenamefont {Biber}\ \emph {et~al.}(2022)\citenamefont {Biber},
  \citenamefont {Brötzner}, \citenamefont {Jäggi}, \citenamefont {Szabo},
  \citenamefont {Pichler}, \citenamefont {Cupak}, \citenamefont {Voith},
  \citenamefont {Cserveny}, \citenamefont {Nenning}, \citenamefont {Mutzke},
  \citenamefont {Moro}, \citenamefont {Primetzhofer}, \citenamefont {Mezger},
  \citenamefont {Galli}, \citenamefont {Wurz},\ and\ \citenamefont
  {Aumayr}}]{biber_sputtering_2022}%
  \BibitemOpen
  \bibfield  {author} {\bibinfo {author} {\bibfnamefont {H.}~\bibnamefont
  {Biber}}, \bibinfo {author} {\bibfnamefont {J.}~\bibnamefont {Brötzner}},
  \bibinfo {author} {\bibfnamefont {N.}~\bibnamefont {Jäggi}}, \bibinfo
  {author} {\bibfnamefont {P.~S.}\ \bibnamefont {Szabo}}, \bibinfo {author}
  {\bibfnamefont {J.}~\bibnamefont {Pichler}}, \bibinfo {author} {\bibfnamefont
  {C.}~\bibnamefont {Cupak}}, \bibinfo {author} {\bibfnamefont
  {C.}~\bibnamefont {Voith}}, \bibinfo {author} {\bibfnamefont
  {B.}~\bibnamefont {Cserveny}}, \bibinfo {author} {\bibfnamefont
  {A.}~\bibnamefont {Nenning}}, \bibinfo {author} {\bibfnamefont
  {A.}~\bibnamefont {Mutzke}}, \bibinfo {author} {\bibfnamefont {M.~V.}\
  \bibnamefont {Moro}}, \bibinfo {author} {\bibfnamefont {D.}~\bibnamefont
  {Primetzhofer}}, \bibinfo {author} {\bibfnamefont {K.}~\bibnamefont
  {Mezger}}, \bibinfo {author} {\bibfnamefont {A.}~\bibnamefont {Galli}},
  \bibinfo {author} {\bibfnamefont {P.}~\bibnamefont {Wurz}},\ and\ \bibinfo
  {author} {\bibfnamefont {F.}~\bibnamefont {Aumayr}},\ }\bibfield  {title}
  {\bibinfo {title} {Sputtering {Behavior} of {Rough}, {Polycrystalline}
  {Mercury} {Analogs}},\ }\href {https://doi.org/10.3847/PSJ/aca402} {\bibfield
   {journal} {\bibinfo  {journal} {The Planetary Science Journal}\ }\textbf
  {\bibinfo {volume} {3}},\ \bibinfo {pages} {271} (\bibinfo {year}
  {2022})}\BibitemShut {NoStop}%
\bibitem [{\citenamefont {Mutzke}\ \emph {et~al.}(2019)\citenamefont {Mutzke},
  \citenamefont {Schneider}, \citenamefont {Eckstein}, \citenamefont {Dohmen},
  \citenamefont {Schmid}, \citenamefont {von Toussaint},\ and\ \citenamefont
  {Bandelow}}]{mutzke_sdtrimsp_2019}%
  \BibitemOpen
  \bibfield  {author} {\bibinfo {author} {\bibfnamefont {A.}~\bibnamefont
  {Mutzke}}, \bibinfo {author} {\bibfnamefont {R.}~\bibnamefont {Schneider}},
  \bibinfo {author} {\bibfnamefont {W.}~\bibnamefont {Eckstein}}, \bibinfo
  {author} {\bibfnamefont {R.}~\bibnamefont {Dohmen}}, \bibinfo {author}
  {\bibfnamefont {K.}~\bibnamefont {Schmid}}, \bibinfo {author} {\bibfnamefont
  {U.}~\bibnamefont {von Toussaint}},\ and\ \bibinfo {author} {\bibfnamefont
  {G.}~\bibnamefont {Bandelow}},\ }\bibfield  {title} {\bibinfo {title}
  {{SDTrimSP} {Version} 5.00},\ }\href@noop {} {\bibfield  {journal} {\bibinfo
  {journal} {IPP-report}\ }\textbf {\bibinfo {volume} {2019-02}},\ \bibinfo
  {pages} {1} (\bibinfo {year} {2019})}\BibitemShut {NoStop}%
\bibitem [{\citenamefont {Hofsäss}\ \emph {et~al.}(2014)\citenamefont
  {Hofsäss}, \citenamefont {Zhang},\ and\ \citenamefont
  {Mutzke}}]{hofsass_simulation_2014}%
  \BibitemOpen
  \bibfield  {author} {\bibinfo {author} {\bibfnamefont {H.}~\bibnamefont
  {Hofsäss}}, \bibinfo {author} {\bibfnamefont {K.}~\bibnamefont {Zhang}},\
  and\ \bibinfo {author} {\bibfnamefont {A.}~\bibnamefont {Mutzke}},\
  }\bibfield  {title} {\bibinfo {title} {Simulation of ion beam sputtering with
  {SDTrimSP}, {TRIDYN} and {SRIM}},\ }\href
  {https://doi.org/10.1016/j.apsusc.2014.03.152} {\bibfield  {journal}
  {\bibinfo  {journal} {Applied Surface Science}\ }\textbf {\bibinfo {volume}
  {310}},\ \bibinfo {pages} {134} (\bibinfo {year} {2014})}\BibitemShut
  {NoStop}%
\bibitem [{\citenamefont {Hofsäss}\ \emph {et~al.}(2023)\citenamefont
  {Hofsäss}, \citenamefont {Junge}, \citenamefont {Kirscht},\ and\
  \citenamefont {Van~Stiphout}}]{hofsass_low_2023}%
  \BibitemOpen
  \bibfield  {author} {\bibinfo {author} {\bibfnamefont {H.}~\bibnamefont
  {Hofsäss}}, \bibinfo {author} {\bibfnamefont {F.}~\bibnamefont {Junge}},
  \bibinfo {author} {\bibfnamefont {P.}~\bibnamefont {Kirscht}},\ and\ \bibinfo
  {author} {\bibfnamefont {K.}~\bibnamefont {Van~Stiphout}},\ }\bibfield
  {title} {\bibinfo {title} {Low energy ion-solid interactions: a quantitative
  experimental verification of binary collision approximation simulations},\
  }\href {https://doi.org/10.1088/2053-1591/ace41c} {\bibfield  {journal}
  {\bibinfo  {journal} {Materials Research Express}\ }\textbf {\bibinfo
  {volume} {10}},\ \bibinfo {pages} {075003} (\bibinfo {year}
  {2023})}\BibitemShut {NoStop}%
\bibitem [{\citenamefont {Primetzhofer}\ \emph {et~al.}(2011)\citenamefont
  {Primetzhofer}, \citenamefont {Markin}, \citenamefont {Efrosinin},
  \citenamefont {Steinbauer}, \citenamefont {Andrzejewski},\ and\ \citenamefont
  {Bauer}}]{primetzhofer_influence_2011}%
  \BibitemOpen
  \bibfield  {author} {\bibinfo {author} {\bibfnamefont {D.}~\bibnamefont
  {Primetzhofer}}, \bibinfo {author} {\bibfnamefont {S.}~\bibnamefont
  {Markin}}, \bibinfo {author} {\bibfnamefont {D.}~\bibnamefont {Efrosinin}},
  \bibinfo {author} {\bibfnamefont {E.}~\bibnamefont {Steinbauer}}, \bibinfo
  {author} {\bibfnamefont {R.}~\bibnamefont {Andrzejewski}},\ and\ \bibinfo
  {author} {\bibfnamefont {P.}~\bibnamefont {Bauer}},\ }\bibfield  {title}
  {\bibinfo {title} {Influence of screening length modification on the
  scattering cross section in {LEIS}},\ }\href
  {https://doi.org/10.1016/j.nimb.2010.11.019} {\bibfield  {journal} {\bibinfo
  {journal} {Nuclear Instruments and Methods in Physics Research Section B:
  Beam Interactions with Materials and Atoms}\ }\textbf {\bibinfo {volume}
  {269}},\ \bibinfo {pages} {1292} (\bibinfo {year} {2011})}\BibitemShut
  {NoStop}%
\bibitem [{\citenamefont {Primetzhofer}\ \emph {et~al.}(2009)\citenamefont
  {Primetzhofer}, \citenamefont {Markin}, \citenamefont {Juaristi},
  \citenamefont {Taglauer},\ and\ \citenamefont
  {Bauer}}]{primetzhofer_leis_2009}%
  \BibitemOpen
  \bibfield  {author} {\bibinfo {author} {\bibfnamefont {D.}~\bibnamefont
  {Primetzhofer}}, \bibinfo {author} {\bibfnamefont {S.}~\bibnamefont
  {Markin}}, \bibinfo {author} {\bibfnamefont {J.}~\bibnamefont {Juaristi}},
  \bibinfo {author} {\bibfnamefont {E.}~\bibnamefont {Taglauer}},\ and\
  \bibinfo {author} {\bibfnamefont {P.}~\bibnamefont {Bauer}},\ }\bibfield
  {title} {\bibinfo {title} {{LEIS}: {A} reliable tool for surface composition
  analysis?},\ }\href {https://doi.org/10.1016/j.nimb.2008.10.050} {\bibfield
  {journal} {\bibinfo  {journal} {Nuclear Instruments and Methods in Physics
  Research Section B: Beam Interactions with Materials and Atoms}\ }\textbf
  {\bibinfo {volume} {267}},\ \bibinfo {pages} {624} (\bibinfo {year}
  {2009})}\BibitemShut {NoStop}%
\bibitem [{\citenamefont {Primetzhofer}\ \emph {et~al.}(2008)\citenamefont
  {Primetzhofer}, \citenamefont {Markin}, \citenamefont {Draxler},
  \citenamefont {Beikler}, \citenamefont {Taglauer},\ and\ \citenamefont
  {Bauer}}]{primetzhofer_strength_2008}%
  \BibitemOpen
  \bibfield  {author} {\bibinfo {author} {\bibfnamefont {D.}~\bibnamefont
  {Primetzhofer}}, \bibinfo {author} {\bibfnamefont {S.}~\bibnamefont
  {Markin}}, \bibinfo {author} {\bibfnamefont {M.}~\bibnamefont {Draxler}},
  \bibinfo {author} {\bibfnamefont {R.}~\bibnamefont {Beikler}}, \bibinfo
  {author} {\bibfnamefont {E.}~\bibnamefont {Taglauer}},\ and\ \bibinfo
  {author} {\bibfnamefont {P.}~\bibnamefont {Bauer}},\ }\bibfield  {title}
  {\bibinfo {title} {Strength of the interatomic potential derived from angular
  scans in {LEIS}},\ }\href {https://doi.org/10.1016/j.susc.2008.07.030}
  {\bibfield  {journal} {\bibinfo  {journal} {Surface Science}\ }\textbf
  {\bibinfo {volume} {602}},\ \bibinfo {pages} {2921} (\bibinfo {year}
  {2008})}\BibitemShut {NoStop}%
\bibitem [{\citenamefont {Lohmann}\ \emph {et~al.}(2023)\citenamefont
  {Lohmann}, \citenamefont {Holeňák}, \citenamefont {Grande},\ and\
  \citenamefont {Primetzhofer}}]{lohmann_trajectory_2023}%
  \BibitemOpen
  \bibfield  {author} {\bibinfo {author} {\bibfnamefont {S.}~\bibnamefont
  {Lohmann}}, \bibinfo {author} {\bibfnamefont {R.}~\bibnamefont {Holeňák}},
  \bibinfo {author} {\bibfnamefont {P.~L.}\ \bibnamefont {Grande}},\ and\
  \bibinfo {author} {\bibfnamefont {D.}~\bibnamefont {Primetzhofer}},\
  }\bibfield  {title} {\bibinfo {title} {Trajectory dependence of electronic
  energy-loss straggling at {keV} ion energies},\ }\href
  {https://doi.org/10.1103/PhysRevB.107.085110} {\bibfield  {journal} {\bibinfo
   {journal} {Physical Review B}\ }\textbf {\bibinfo {volume} {107}},\ \bibinfo
  {pages} {085110} (\bibinfo {year} {2023})}\BibitemShut {NoStop}%
\bibitem [{\citenamefont {Průša}\ \emph {et~al.}(2024)\citenamefont
  {Průša}, \citenamefont {Linford}, \citenamefont {Vaníčková},
  \citenamefont {Bábík}, \citenamefont {Pinder}, \citenamefont {Šikola},\
  and\ \citenamefont {Brongersma}}]{prusa_practical_2024}%
  \BibitemOpen
  \bibfield  {author} {\bibinfo {author} {\bibfnamefont {S.}~\bibnamefont
  {Průša}}, \bibinfo {author} {\bibfnamefont {M.~R.}\ \bibnamefont
  {Linford}}, \bibinfo {author} {\bibfnamefont {E.}~\bibnamefont
  {Vaníčková}}, \bibinfo {author} {\bibfnamefont {P.}~\bibnamefont
  {Bábík}}, \bibinfo {author} {\bibfnamefont {J.~W.}\ \bibnamefont {Pinder}},
  \bibinfo {author} {\bibfnamefont {T.}~\bibnamefont {Šikola}},\ and\ \bibinfo
  {author} {\bibfnamefont {H.~H.}\ \bibnamefont {Brongersma}},\ }\bibfield
  {title} {\bibinfo {title} {A practical guide to interpreting low energy ion
  scattering ({LEIS}) spectra},\ }\href
  {https://doi.org/10.1016/j.apsusc.2023.158793} {\bibfield  {journal}
  {\bibinfo  {journal} {Applied Surface Science}\ }\textbf {\bibinfo {volume}
  {657}},\ \bibinfo {pages} {158793} (\bibinfo {year} {2024})}\BibitemShut
  {NoStop}%
\bibitem [{\citenamefont {Téllez}\ \emph {et~al.}(2014)\citenamefont
  {Téllez}, \citenamefont {Aguadero}, \citenamefont {Druce}, \citenamefont
  {Burriel}, \citenamefont {Fearn}, \citenamefont {Ishihara}, \citenamefont
  {McPhail},\ and\ \citenamefont {Kilner}}]{tellez_new_2014}%
  \BibitemOpen
  \bibfield  {author} {\bibinfo {author} {\bibfnamefont {H.}~\bibnamefont
  {Téllez}}, \bibinfo {author} {\bibfnamefont {A.}~\bibnamefont {Aguadero}},
  \bibinfo {author} {\bibfnamefont {J.}~\bibnamefont {Druce}}, \bibinfo
  {author} {\bibfnamefont {M.}~\bibnamefont {Burriel}}, \bibinfo {author}
  {\bibfnamefont {S.}~\bibnamefont {Fearn}}, \bibinfo {author} {\bibfnamefont
  {T.}~\bibnamefont {Ishihara}}, \bibinfo {author} {\bibfnamefont {D.~S.}\
  \bibnamefont {McPhail}},\ and\ \bibinfo {author} {\bibfnamefont {J.~A.}\
  \bibnamefont {Kilner}},\ }\bibfield  {title} {\bibinfo {title} {New
  perspectives in the surface analysis of energy materials by combined
  time-of-flight secondary ion mass spectrometry ({ToF}-{SIMS}) and high
  sensitivity low-energy ion scattering ({HS}-{LEIS})},\ }\href
  {https://doi.org/10.1039/C3JA50292A} {\bibfield  {journal} {\bibinfo
  {journal} {Journal of Analytical Atomic Spectrometry}\ }\textbf {\bibinfo
  {volume} {29}},\ \bibinfo {pages} {1361} (\bibinfo {year}
  {2014})}\BibitemShut {NoStop}%
\bibitem [{\citenamefont {Avval}\ \emph {et~al.}(2019)\citenamefont {Avval},
  \citenamefont {Cushman}, \citenamefont {Brüner}, \citenamefont {Grehl},
  \citenamefont {Brongersma},\ and\ \citenamefont
  {Linford}}]{avval_calcium_2019}%
  \BibitemOpen
  \bibfield  {author} {\bibinfo {author} {\bibfnamefont {T.~G.}\ \bibnamefont
  {Avval}}, \bibinfo {author} {\bibfnamefont {C.~V.}\ \bibnamefont {Cushman}},
  \bibinfo {author} {\bibfnamefont {P.}~\bibnamefont {Brüner}}, \bibinfo
  {author} {\bibfnamefont {T.}~\bibnamefont {Grehl}}, \bibinfo {author}
  {\bibfnamefont {H.~H.}\ \bibnamefont {Brongersma}},\ and\ \bibinfo {author}
  {\bibfnamefont {M.~R.}\ \bibnamefont {Linford}},\ }\bibfield  {title}
  {\bibinfo {title} {Calcium fluoride and gold reference by high
  sensitivity-low energy ion scattering},\ }\href
  {https://doi.org/10.1116/1.5115065} {\bibfield  {journal} {\bibinfo
  {journal} {Surface Science Spectra}\ }\textbf {\bibinfo {volume} {26}},\
  \bibinfo {pages} {024201} (\bibinfo {year} {2019})}\BibitemShut {NoStop}%
\bibitem [{\citenamefont {Cushman}\ \emph {et~al.}(2018)\citenamefont
  {Cushman}, \citenamefont {Brüner}, \citenamefont {Zakel}, \citenamefont
  {Dahlquist}, \citenamefont {Sturgell}, \citenamefont {Grehl}, \citenamefont
  {Lunt}, \citenamefont {Banerjee}, \citenamefont {Smith},\ and\ \citenamefont
  {Linford}}]{cushman_low_2018}%
  \BibitemOpen
  \bibfield  {author} {\bibinfo {author} {\bibfnamefont {C.~V.}\ \bibnamefont
  {Cushman}}, \bibinfo {author} {\bibfnamefont {P.}~\bibnamefont {Brüner}},
  \bibinfo {author} {\bibfnamefont {J.}~\bibnamefont {Zakel}}, \bibinfo
  {author} {\bibfnamefont {C.}~\bibnamefont {Dahlquist}}, \bibinfo {author}
  {\bibfnamefont {B.}~\bibnamefont {Sturgell}}, \bibinfo {author}
  {\bibfnamefont {T.}~\bibnamefont {Grehl}}, \bibinfo {author} {\bibfnamefont
  {B.~M.}\ \bibnamefont {Lunt}}, \bibinfo {author} {\bibfnamefont
  {J.}~\bibnamefont {Banerjee}}, \bibinfo {author} {\bibfnamefont {N.~J.}\
  \bibnamefont {Smith}},\ and\ \bibinfo {author} {\bibfnamefont {M.~R.}\
  \bibnamefont {Linford}},\ }\bibfield  {title} {\bibinfo {title} {Low energy
  ion scattering ({LEIS}) of as-formed and chemically modified display glass
  and peak-fitting of the {Al}/{Si} {LEIS} peak envelope},\ }\href
  {https://doi.org/10.1016/j.apsusc.2018.04.127} {\bibfield  {journal}
  {\bibinfo  {journal} {Applied Surface Science}\ }\textbf {\bibinfo {volume}
  {455}},\ \bibinfo {pages} {18} (\bibinfo {year} {2018})}\BibitemShut
  {NoStop}%
\bibitem [{\citenamefont {Gasteiger}\ \emph {et~al.}(1993)\citenamefont
  {Gasteiger}, \citenamefont {Ross},\ and\ \citenamefont
  {Cairns}}]{gasteiger_leis_1993}%
  \BibitemOpen
  \bibfield  {author} {\bibinfo {author} {\bibfnamefont {H.~A.}\ \bibnamefont
  {Gasteiger}}, \bibinfo {author} {\bibfnamefont {P.~N.}\ \bibnamefont
  {Ross}},\ and\ \bibinfo {author} {\bibfnamefont {E.~J.}\ \bibnamefont
  {Cairns}},\ }\bibfield  {title} {\bibinfo {title} {{LEIS} and {AES} on
  sputtered and annealed polycrystalline {Pt}-{Ru} bulk alloys},\ }\href
  {https://doi.org/10.1016/0039-6028(93)90244-E} {\bibfield  {journal}
  {\bibinfo  {journal} {Surface Science}\ }\textbf {\bibinfo {volume} {293}},\
  \bibinfo {pages} {67} (\bibinfo {year} {1993})}\BibitemShut {NoStop}%
\bibitem [{\citenamefont {Goebl}\ \emph {et~al.}(2015)\citenamefont {Goebl},
  \citenamefont {Bruckner}, \citenamefont {Roth}, \citenamefont {Ahamer},\ and\
  \citenamefont {Bauer}}]{goebl_low-energy_2015}%
  \BibitemOpen
  \bibfield  {author} {\bibinfo {author} {\bibfnamefont {D.}~\bibnamefont
  {Goebl}}, \bibinfo {author} {\bibfnamefont {B.}~\bibnamefont {Bruckner}},
  \bibinfo {author} {\bibfnamefont {D.}~\bibnamefont {Roth}}, \bibinfo {author}
  {\bibfnamefont {C.}~\bibnamefont {Ahamer}},\ and\ \bibinfo {author}
  {\bibfnamefont {P.}~\bibnamefont {Bauer}},\ }\bibfield  {title} {\bibinfo
  {title} {Low-energy ion scattering: {A} quantitative method?},\ }\href
  {https://doi.org/10.1016/j.nimb.2014.11.030} {\bibfield  {journal} {\bibinfo
  {journal} {Nuclear Instruments and Methods in Physics Research Section B:
  Beam Interactions with Materials and Atoms}\ }\textbf {\bibinfo {volume}
  {354}},\ \bibinfo {pages} {3} (\bibinfo {year} {2015})}\BibitemShut {NoStop}%
\bibitem [{\citenamefont {Creutzburg}\ \emph {et~al.}(2021)\citenamefont
  {Creutzburg}, \citenamefont {Niggas}, \citenamefont {Weichselbaum},
  \citenamefont {Grande}, \citenamefont {Aumayr},\ and\ \citenamefont
  {Wilhelm}}]{creutzburg_angle-dependent_2021}%
  \BibitemOpen
  \bibfield  {author} {\bibinfo {author} {\bibfnamefont {S.}~\bibnamefont
  {Creutzburg}}, \bibinfo {author} {\bibfnamefont {A.}~\bibnamefont {Niggas}},
  \bibinfo {author} {\bibfnamefont {D.}~\bibnamefont {Weichselbaum}}, \bibinfo
  {author} {\bibfnamefont {P.~L.}\ \bibnamefont {Grande}}, \bibinfo {author}
  {\bibfnamefont {F.}~\bibnamefont {Aumayr}},\ and\ \bibinfo {author}
  {\bibfnamefont {R.~A.}\ \bibnamefont {Wilhelm}},\ }\bibfield  {title}
  {\bibinfo {title} {Angle-dependent charge exchange and energy loss of slow
  highly charged ions in freestanding graphene},\ }\href
  {https://doi.org/10.1103/PhysRevA.104.042806} {\bibfield  {journal} {\bibinfo
   {journal} {Physical Review A}\ }\textbf {\bibinfo {volume} {104}},\ \bibinfo
  {pages} {042806} (\bibinfo {year} {2021})}\BibitemShut {NoStop}%
\bibitem [{\citenamefont {Wilhelm}\ and\ \citenamefont
  {Grande}(2019)}]{wilhelm_unraveling_2019}%
  \BibitemOpen
  \bibfield  {author} {\bibinfo {author} {\bibfnamefont {R.}~\bibnamefont
  {Wilhelm}}\ and\ \bibinfo {author} {\bibfnamefont {P.}~\bibnamefont
  {Grande}},\ }\bibfield  {title} {\bibinfo {title} {Unraveling energy loss
  processes of low energy heavy ions in {2D} materials},\ }\href
  {https://doi.org/10.1038/s42005-019-0188-7} {\bibfield  {journal} {\bibinfo
  {journal} {Communications Physics}\ }\textbf {\bibinfo {volume} {2}},\
  \bibinfo {pages} {89} (\bibinfo {year} {2019})}\BibitemShut {NoStop}%
\bibitem [{\citenamefont {Arezki}\ \emph {et~al.}(1998)\citenamefont {Arezki},
  \citenamefont {Boudouma}, \citenamefont {Benoit-Cattin}, \citenamefont
  {Chami}, \citenamefont {Benazeth}, \citenamefont {Khalal},\ and\
  \citenamefont {Boudjema}}]{arezki_angular_1998}%
  \BibitemOpen
  \bibfield  {author} {\bibinfo {author} {\bibfnamefont {B.}~\bibnamefont
  {Arezki}}, \bibinfo {author} {\bibfnamefont {Y.}~\bibnamefont {Boudouma}},
  \bibinfo {author} {\bibfnamefont {P.}~\bibnamefont {Benoit-Cattin}}, \bibinfo
  {author} {\bibfnamefont {A.~C.}\ \bibnamefont {Chami}}, \bibinfo {author}
  {\bibfnamefont {C.}~\bibnamefont {Benazeth}}, \bibinfo {author}
  {\bibfnamefont {K.}~\bibnamefont {Khalal}},\ and\ \bibinfo {author}
  {\bibfnamefont {M.}~\bibnamefont {Boudjema}},\ }\bibfield  {title} {\bibinfo
  {title} {Angular, energy and charge distribution in the scattering of
  low-energy helium ions by an amorphous silicon surface},\ }\href
  {https://doi.org/10.1088/0953-8984/10/4/004} {\bibfield  {journal} {\bibinfo
  {journal} {Journal of Physics: Condensed Matter}\ }\textbf {\bibinfo {volume}
  {10}},\ \bibinfo {pages} {741} (\bibinfo {year} {1998})}\BibitemShut
  {NoStop}%
\bibitem [{\citenamefont {Sturm}\ \emph {et~al.}(2023)\citenamefont {Sturm},
  \citenamefont {Lokhorst}, \citenamefont {Zameshin},\ and\ \citenamefont
  {Ackermann}}]{sturm_charge_2023}%
  \BibitemOpen
  \bibfield  {author} {\bibinfo {author} {\bibfnamefont {J.}~\bibnamefont
  {Sturm}}, \bibinfo {author} {\bibfnamefont {H.}~\bibnamefont {Lokhorst}},
  \bibinfo {author} {\bibfnamefont {A.}~\bibnamefont {Zameshin}},\ and\
  \bibinfo {author} {\bibfnamefont {M.}~\bibnamefont {Ackermann}},\ }\bibfield
  {title} {\bibinfo {title} {Charge exchange between {He}+ ions and solid
  targets: {The} dependence on target electronic structure revisited},\ }\href
  {https://doi.org/10.1016/j.nimb.2023.02.029} {\bibfield  {journal} {\bibinfo
  {journal} {Nuclear Instruments and Methods in Physics Research Section B:
  Beam Interactions with Materials and Atoms}\ }\textbf {\bibinfo {volume}
  {538}},\ \bibinfo {pages} {47} (\bibinfo {year} {2023})}\BibitemShut
  {NoStop}%
\bibitem [{\citenamefont {Arnau}\ \emph {et~al.}(1995)\citenamefont {Arnau},
  \citenamefont {Köhrbrück}, \citenamefont {Grether}, \citenamefont
  {Spieler},\ and\ \citenamefont {Stolterfoht}}]{arnau_molecular-orbital_1995}%
  \BibitemOpen
  \bibfield  {author} {\bibinfo {author} {\bibfnamefont {A.}~\bibnamefont
  {Arnau}}, \bibinfo {author} {\bibfnamefont {R.}~\bibnamefont {Köhrbrück}},
  \bibinfo {author} {\bibfnamefont {M.}~\bibnamefont {Grether}}, \bibinfo
  {author} {\bibfnamefont {A.}~\bibnamefont {Spieler}},\ and\ \bibinfo {author}
  {\bibfnamefont {N.}~\bibnamefont {Stolterfoht}},\ }\bibfield  {title}
  {\bibinfo {title} {Molecular-orbital model for slow hollow atoms colliding
  with atoms in a solid},\ }\href {https://doi.org/10.1103/PhysRevA.51.R3399}
  {\bibfield  {journal} {\bibinfo  {journal} {Physical Review A}\ }\textbf
  {\bibinfo {volume} {51}},\ \bibinfo {pages} {R3399} (\bibinfo {year}
  {1995})}\BibitemShut {NoStop}%
\bibitem [{\citenamefont {Fano}\ and\ \citenamefont
  {Lichten}(1965)}]{fano_interpretation_1965}%
  \BibitemOpen
  \bibfield  {author} {\bibinfo {author} {\bibfnamefont {U.}~\bibnamefont
  {Fano}}\ and\ \bibinfo {author} {\bibfnamefont {W.}~\bibnamefont {Lichten}},\
  }\bibfield  {title} {\bibinfo {title} {Interpretation of {Ar}+ - {Ar}
  {Collisions} at 50 {KeV}},\ }\href
  {https://doi.org/10.1103/PhysRevLett.14.627} {\bibfield  {journal} {\bibinfo
  {journal} {Physical Review Letters}\ }\textbf {\bibinfo {volume} {14}},\
  \bibinfo {pages} {627} (\bibinfo {year} {1965})}\BibitemShut {NoStop}%
\bibitem [{\citenamefont {Sasaki}\ \emph {et~al.}(2002)\citenamefont {Sasaki},
  \citenamefont {Scanlon}, \citenamefont {Ermolov},\ and\ \citenamefont
  {Brongersma}}]{sasaki_neutralization_2002}%
  \BibitemOpen
  \bibfield  {author} {\bibinfo {author} {\bibfnamefont {M.}~\bibnamefont
  {Sasaki}}, \bibinfo {author} {\bibfnamefont {P.~J.}\ \bibnamefont {Scanlon}},
  \bibinfo {author} {\bibfnamefont {S.}~\bibnamefont {Ermolov}},\ and\ \bibinfo
  {author} {\bibfnamefont {H.~H.}\ \bibnamefont {Brongersma}},\ }\bibfield
  {title} {\bibinfo {title} {Neutralization of {He} ions scattered from {Ca}
  surface},\ }\href {https://doi.org/10.1016/S0168-583X(01)01207-1} {\bibfield
  {journal} {\bibinfo  {journal} {Nuclear Instruments and Methods in Physics
  Research Section B: Beam Interactions with Materials and Atoms}\ }\textbf
  {\bibinfo {volume} {190}},\ \bibinfo {pages} {127} (\bibinfo {year}
  {2002})}\BibitemShut {NoStop}%
\bibitem [{\citenamefont {Verbist}\ \emph {et~al.}(1992)\citenamefont
  {Verbist}, \citenamefont {Brongersma},\ and\ \citenamefont
  {Devreese}}]{verbist_calculation_1992}%
  \BibitemOpen
  \bibfield  {author} {\bibinfo {author} {\bibfnamefont {G.}~\bibnamefont
  {Verbist}}, \bibinfo {author} {\bibfnamefont {H.~H.}\ \bibnamefont
  {Brongersma}},\ and\ \bibinfo {author} {\bibfnamefont {J.~T.}\ \bibnamefont
  {Devreese}},\ }\bibfield  {title} {\bibinfo {title} {The calculation of ion
  fractions in {LEIS}},\ }\href {https://doi.org/10.1016/0168-583X(92)95535-Y}
  {\bibfield  {journal} {\bibinfo  {journal} {Nuclear Instruments and Methods
  in Physics Research Section B: Beam Interactions with Materials and Atoms}\
  }\textbf {\bibinfo {volume} {64}},\ \bibinfo {pages} {572} (\bibinfo {year}
  {1992})}\BibitemShut {NoStop}%
\bibitem [{\citenamefont {Bruckner}\ \emph {et~al.}(2020)\citenamefont
  {Bruckner}, \citenamefont {Bauer},\ and\ \citenamefont
  {Primetzhofer}}]{bruckner_neutralization_2020}%
  \BibitemOpen
  \bibfield  {author} {\bibinfo {author} {\bibfnamefont {B.}~\bibnamefont
  {Bruckner}}, \bibinfo {author} {\bibfnamefont {P.}~\bibnamefont {Bauer}},\
  and\ \bibinfo {author} {\bibfnamefont {D.}~\bibnamefont {Primetzhofer}},\
  }\bibfield  {title} {\bibinfo {title} {Neutralization of slow helium ions
  scattered from single crystalline aluminum and tantalum surfaces and their
  oxides},\ }\href {https://doi.org/10.1016/j.susc.2019.121491} {\bibfield
  {journal} {\bibinfo  {journal} {Surface Science}\ }\textbf {\bibinfo {volume}
  {691}},\ \bibinfo {pages} {121491} (\bibinfo {year} {2020})}\BibitemShut
  {NoStop}%
\bibitem [{\citenamefont {Bhardwaj}\ \emph {et~al.}(2015)\citenamefont
  {Bhardwaj}, \citenamefont {Dhanya}, \citenamefont {Alok}, \citenamefont
  {Barabash}, \citenamefont {Wieser}, \citenamefont {Futaana}, \citenamefont
  {Wurz}, \citenamefont {Vorburger}, \citenamefont {Holmström}, \citenamefont
  {Lue}, \citenamefont {Harada},\ and\ \citenamefont
  {Asamura}}]{bhardwaj_new_2015}%
  \BibitemOpen
  \bibfield  {author} {\bibinfo {author} {\bibfnamefont {A.}~\bibnamefont
  {Bhardwaj}}, \bibinfo {author} {\bibfnamefont {M.~B.}\ \bibnamefont
  {Dhanya}}, \bibinfo {author} {\bibfnamefont {A.}~\bibnamefont {Alok}},
  \bibinfo {author} {\bibfnamefont {S.}~\bibnamefont {Barabash}}, \bibinfo
  {author} {\bibfnamefont {M.}~\bibnamefont {Wieser}}, \bibinfo {author}
  {\bibfnamefont {Y.}~\bibnamefont {Futaana}}, \bibinfo {author} {\bibfnamefont
  {P.}~\bibnamefont {Wurz}}, \bibinfo {author} {\bibfnamefont {A.}~\bibnamefont
  {Vorburger}}, \bibinfo {author} {\bibfnamefont {M.}~\bibnamefont
  {Holmström}}, \bibinfo {author} {\bibfnamefont {C.}~\bibnamefont {Lue}},
  \bibinfo {author} {\bibfnamefont {Y.}~\bibnamefont {Harada}},\ and\ \bibinfo
  {author} {\bibfnamefont {K.}~\bibnamefont {Asamura}},\ }\bibfield  {title}
  {\bibinfo {title} {A new view on the solar wind interaction with the
  {Moon}},\ }\href {https://doi.org/10.1186/s40562-015-0027-y} {\bibfield
  {journal} {\bibinfo  {journal} {Geoscience Letters}\ }\textbf {\bibinfo
  {volume} {2}},\ \bibinfo {pages} {10} (\bibinfo {year} {2015})}\BibitemShut
  {NoStop}%
\bibitem [{\citenamefont {Lue}\ \emph {et~al.}(2018)\citenamefont {Lue},
  \citenamefont {Halekas}, \citenamefont {Poppe},\ and\ \citenamefont
  {McFadden}}]{lue_artemis_2018}%
  \BibitemOpen
  \bibfield  {author} {\bibinfo {author} {\bibfnamefont {C.}~\bibnamefont
  {Lue}}, \bibinfo {author} {\bibfnamefont {J.~S.}\ \bibnamefont {Halekas}},
  \bibinfo {author} {\bibfnamefont {A.~R.}\ \bibnamefont {Poppe}},\ and\
  \bibinfo {author} {\bibfnamefont {J.~P.}\ \bibnamefont {McFadden}},\
  }\bibfield  {title} {\bibinfo {title} {{ARTEMIS} {Observations} of {Solar}
  {Wind} {Proton} {Scattering} off the {Lunar} {Surface}},\ }\href
  {https://doi.org/10.1029/2018JA025486} {\bibfield  {journal} {\bibinfo
  {journal} {Journal of Geophysical Research: Space Physics}\ }\textbf
  {\bibinfo {volume} {123}},\ \bibinfo {pages} {5289} (\bibinfo {year}
  {2018})}\BibitemShut {NoStop}%
\bibitem [{\citenamefont {Pieters}\ and\ \citenamefont
  {Noble}(2016)}]{pieters_space_2016}%
  \BibitemOpen
  \bibfield  {author} {\bibinfo {author} {\bibfnamefont {C.~M.}\ \bibnamefont
  {Pieters}}\ and\ \bibinfo {author} {\bibfnamefont {S.~K.}\ \bibnamefont
  {Noble}},\ }\bibfield  {title} {\bibinfo {title} {Space weathering on airless
  bodies: {SPACE} {WEATHERING} {ON} {AIRLESS} {BODIES}},\ }\href
  {https://doi.org/10.1002/2016JE005128} {\bibfield  {journal} {\bibinfo
  {journal} {Journal of Geophysical Research: Planets}\ }\textbf {\bibinfo
  {volume} {121}},\ \bibinfo {pages} {1865} (\bibinfo {year}
  {2016})}\BibitemShut {NoStop}%
\bibitem [{\citenamefont {Szabo}\ \emph {et~al.}(2020)\citenamefont {Szabo},
  \citenamefont {Biber}, \citenamefont {Jäggi}, \citenamefont {Brenner},
  \citenamefont {Weichselbaum}, \citenamefont {Niggas}, \citenamefont
  {Stadlmayr}, \citenamefont {Primetzhofer}, \citenamefont {Nenning},
  \citenamefont {Mutzke}, \citenamefont {Sauer}, \citenamefont {Fleig},
  \citenamefont {Foelske-Schmitz}, \citenamefont {Mezger}, \citenamefont
  {Lammer}, \citenamefont {Galli}, \citenamefont {Wurz},\ and\ \citenamefont
  {Aumayr}}]{szabo_dynamic_2020}%
  \BibitemOpen
  \bibfield  {author} {\bibinfo {author} {\bibfnamefont {P.~S.}\ \bibnamefont
  {Szabo}}, \bibinfo {author} {\bibfnamefont {H.}~\bibnamefont {Biber}},
  \bibinfo {author} {\bibfnamefont {N.}~\bibnamefont {Jäggi}}, \bibinfo
  {author} {\bibfnamefont {M.}~\bibnamefont {Brenner}}, \bibinfo {author}
  {\bibfnamefont {D.}~\bibnamefont {Weichselbaum}}, \bibinfo {author}
  {\bibfnamefont {A.}~\bibnamefont {Niggas}}, \bibinfo {author} {\bibfnamefont
  {R.}~\bibnamefont {Stadlmayr}}, \bibinfo {author} {\bibfnamefont
  {D.}~\bibnamefont {Primetzhofer}}, \bibinfo {author} {\bibfnamefont
  {A.}~\bibnamefont {Nenning}}, \bibinfo {author} {\bibfnamefont
  {A.}~\bibnamefont {Mutzke}}, \bibinfo {author} {\bibfnamefont
  {M.}~\bibnamefont {Sauer}}, \bibinfo {author} {\bibfnamefont
  {J.}~\bibnamefont {Fleig}}, \bibinfo {author} {\bibfnamefont
  {A.}~\bibnamefont {Foelske-Schmitz}}, \bibinfo {author} {\bibfnamefont
  {K.}~\bibnamefont {Mezger}}, \bibinfo {author} {\bibfnamefont
  {H.}~\bibnamefont {Lammer}}, \bibinfo {author} {\bibfnamefont
  {A.}~\bibnamefont {Galli}}, \bibinfo {author} {\bibfnamefont
  {P.}~\bibnamefont {Wurz}},\ and\ \bibinfo {author} {\bibfnamefont
  {F.}~\bibnamefont {Aumayr}},\ }\bibfield  {title} {\bibinfo {title} {Dynamic
  {Potential} {Sputtering} of {Lunar} {Analog} {Material} by {Solar} {Wind}
  {Ions}},\ }\href {https://doi.org/10.3847/1538-4357/ab7008} {\bibfield
  {journal} {\bibinfo  {journal} {The Astrophysical Journal}\ }\textbf
  {\bibinfo {volume} {891}},\ \bibinfo {pages} {100} (\bibinfo {year}
  {2020})}\BibitemShut {NoStop}%
\bibitem [{\citenamefont {Newville}\ \emph {et~al.}(2025)\citenamefont
  {Newville}, \citenamefont {Otten}, \citenamefont {Nelson}, \citenamefont
  {Stensitzki}, \citenamefont {Ingargiola}, \citenamefont {Allan},
  \citenamefont {Fox}, \citenamefont {Carter},\ and\ \citenamefont
  {Rawlik}}]{newville_lmfit_2025}%
  \BibitemOpen
  \bibfield  {author} {\bibinfo {author} {\bibfnamefont {M.}~\bibnamefont
  {Newville}}, \bibinfo {author} {\bibfnamefont {R.}~\bibnamefont {Otten}},
  \bibinfo {author} {\bibfnamefont {A.}~\bibnamefont {Nelson}}, \bibinfo
  {author} {\bibfnamefont {T.}~\bibnamefont {Stensitzki}}, \bibinfo {author}
  {\bibfnamefont {A.}~\bibnamefont {Ingargiola}}, \bibinfo {author}
  {\bibfnamefont {D.}~\bibnamefont {Allan}}, \bibinfo {author} {\bibfnamefont
  {A.}~\bibnamefont {Fox}}, \bibinfo {author} {\bibfnamefont {F.}~\bibnamefont
  {Carter}},\ and\ \bibinfo {author} {\bibfnamefont {M.}~\bibnamefont
  {Rawlik}},\ }\href {https://doi.org/10.5281/zenodo.16175987} {\bibinfo
  {title} {{LMFIT}: {Non}-{Linear} {Least}-{Squares} {Minimization} and
  {Curve}-{Fitting} for {Python}}} (\bibinfo {year} {2025})\BibitemShut
  {NoStop}%
\end{thebibliography}
\end{document}